\documentclass[12pt]{article}
\usepackage{epsf}
\title{
Glueball spectrum  and the  Pomeron  in the Wilson
loop
approach}
\author{ A.B.Kaidalov and Yu.A.Simonov\\ State Recearch
Center\\Institute of Theoretical and Experimental Physics, \\ Moscow,
Russia}
 \date{}
  \newcommand{\be}{\begin{equation}}
 \newcommand{\ee}{\end{equation}}  
\def\fun#1#2{\lower3.6pt\vbox{\baselineskip0pt\lineskip.9pt
\ialign{$\mathsurround=0pt#1\hfil
##\hfil$\crcr#2\crcr\sim\crcr}}}

\newcommand{\ver}{\mbox{\boldmath${\rm r}$}}

\newcommand{\vep}{\mbox{\boldmath${\rm p}$}}

\newcommand{\veL}{\mbox{\boldmath${\rm L}$}}
\newcommand{\veS}{\mbox{\boldmath${\rm S}$}}
\newcommand{\veB}{\mbox{\boldmath${\rm B}$}}
\newcommand{\veE}{\mbox{\boldmath${\rm E}$}}
\newcommand{\ven}{\mbox{\boldmath${\rm n}$}}
\newcommand{\vexi}{\mbox{\boldmath${\rm \xi}$}}
\newcommand{\veet}{\mbox{\boldmath${\rm \eta}$}}

\begin{document}
\maketitle

 \begin{abstract}
 Using a nonperturbative method based on asymptotic behaviour of
 Wilson loops  we calculate masses of glueballs and corresponding
 Regge--trajectories.  The only input is string tension
 fixed by meson Regge slope, while perturbative
 contributions to spin splittings are defined by standard
 $\alpha_s$ values.
 The masses of  lowest glueball states are in  a perfect agreement
 with lattice results. The leading glueball trajectory which is
 associated with Pomeron is discussed in details and its mixing with
 $f$ and $f'$ trajectories is taken  into account.
   \end{abstract}

 \section{Introduction}

The problem of existence of glueballs is one of the most interesting
in QCD.  Lattice calculations give definite predictions for a
 spectrum of such states [1-4], but experimental evidences are not
 conclusive [5,6]. Mixing between gluons and $q\bar q$-pairs makes a
 separation of glueballs a difficult problem. A theoretical study of
 glueballs in QCD started in [7-10]
 and is closely related to the
 problem of  Pomeron -- leading Regge pole, which determines the
 asymptotic behaviour of scattering amplitudes at very high energies.
 It is usually assumed that  Pomeron in QCD is mostly gluonic
 object [11] and glueball resonances with vacuum quantum numbers and
 largest spins belong to this trajectory. Another interesting
 hypothetical Regge singularity is the "odderon", which has negative
 signature and $C$-parity and can be built out of at least 3 gluons.
 Most studies of the Pomeron and odderon singularities in QCD are
 based on applications of the perturbation theory [12].

 In this paper we will address both the problem of spectra of glueballs and
 of the Pomeron (odderon) singularity using the method of Wilson-loop path
 integrals developed in papers [13-15]. The method is based on the
 assumption of the area law for Wilson loops at large distances in
 QCD, which is equivalent to the condition of confinement of quarks
 and gluons. It has been first applied to calculation of spectra of
 $q\bar   q$--states  in [16], baryons in [17] and glueballs in
 [18]. In this latter  calculation rotation of the string between
 quarks or gluons was not taken into account, which lead to some
 distortion of mass spectra, notably the Regge slope was $1/8\sigma$
 instead of string slope $1/2\pi\sigma$. In this paper we will use
 for glueballs more accurate calculational method developed in
 [14,15] for $q\bar q$--state, which yields the correct Regge slope
 (see [9] for numerical data and discussion). We also make a
 detailed study of possible corrections to large distance string
 dynamics due to small distance perturbative gluon  exchanges (PGE)
 and will demonstrate that their influence on the mass spectrum of
 glueball states is rather small and can be computed as a correction.
 This is in contrast to the glueball spectrum in [8] where PGE in the
 form of the adjoint Coulomb potential was assumed  as in many
 other papers on the subject. Instead we argue below in the paper,
 that PGE sums up to another series, the BFKL  ladder [12], where
 loop corrections strongly suppress the final result , so that PGE
 can be disregarded in the  first approximation.  Our predictions for
   masses of lowest spin-averaged glueball
 states in units of $\sqrt{\sigma}$ are in
 a perfect agreement with results of recent lattice calculations
 [1-3].  In addition spin-orbit and spin-spin interactions are also
 calculated and found in a good agreement with lattice data.

 The leading glueball Regge-trajectory is calculated in the positive
$ t$ region and is extrapolated to the scattering region of $t\le 0$.
 The importance of mixing among this trajectory and $q\bar
 q$-trajectories ($f,f'$) is emphasized; calculation of these mixing
 effects yields the leading Pomeron trajectory with
 $\alpha_P(0)>1$ $(\alpha_P(0)=1.1\div 1.2)$ in accord with
 experimental observations [20]. An interesting pattern of 3
 colliding vacuum trajectories in the region $t>0$ is observed,
 which can be important for decay properties of resonances
 situated at these trajectories.

   The paper is organized as follows.
  The field
 operators creating glueball states in general case and in the
 background field method are introduced in Section 2 and  presented
 in Appendix 1.  The general formalism of the Wilson loop path
 integrals and the resulting relativistic Hamiltonian is discussed in
 Section 2 and Appendix 2.  The spectrum of spin-averaged glueball
 states following from this Hamiltonian is obtained in  Section 2 and
 compared with corresponding lattice results.  Spin splittings of
 glueball masses from both nonperturbative and perturbative parts are
 considered in  Section 3 and Appendix 3 and 4.  Resulting glueball
 spectrum is compared in the Section with lattice calculations.
 The influence of PGE on glueball masses
 and Regge - trajectories is discussed in Section 4. It is pointed
 out that effects of small-distances on glueball spectrum are small.
 Three--gluon glueballs are considered in Section 5. It is shown that
 the lowest 3g states have rather large mass $M_{3g}\approx 3.4 GeV$.
 Relation between glueball Regge trajectories and  the vacuum Pomeron
 trajectory is discussed in Section 6.  In this Section and Appendix
 5  effects of mixing between gluonic and $q\bar
 q$-Regge-trajectories are investigated.

 Possible implications and improvements of the results of the paper
  are discussed in Conclusions.

   \section{General Formalism}

 Following [13,21], we separate gluonic fields $A_\mu$ into
 nonperturbative background $B_\mu$ and perturbative gluons $a_\mu$,
 $A_\mu=B_\mu+a_\mu$
  \footnote{Note that the background formalism
 exploiting the 'tHooft identity [21] allows us to avoid the
 double--counting problem and the principle of separation is unimportant
 provided background $B_\mu$ is characterized by the input string
 tension $\sigma$ and perturbation is in the known $\alpha_s$
 constant}, and consider two-gluon glueballs, described by the
 Green's functions
 $$
 G_{\mu\nu, \mu'\nu'}(x,y|x',y')=
 \langle\Psi^{(in)}(x,y) \Psi^{(out)}(x',y')\rangle_{a,B}=
 $$
 \be
 \langle
 \Gamma^{(in)} G_{\mu\mu'}(x, x') G_{\nu\nu'}(y,y')
 \Gamma^{(out)}\rangle_B +{\rm
 perm}
  \ee
   where $\Psi^{(in)}(\Psi^{(out)})$ are glueball
 operators in the initial (final) state made of gluon
 field $a_\mu$ and $B_\mu$ (see Appendix 1 for explicit
 form of $\Psi^{(in),(out)}$ in lowest states); $G_{\mu\mu'}$ is the
 gluon Green's function of field $a_\mu$ in the background field
 $B_\mu$, namely
  \be
   G_{\mu\nu}(x,y) = \langle x|(-\hat
 D^2\delta_{\mu\nu} -2ig \hat F_{\mu\nu})^{-1}|y\rangle_B
  \ee
   where
 $\hat D_{\mu}=\partial_\mu-ig \hat B_\mu, ~\hat F_{\mu\nu}$ is the
 field strength of the field $\hat B_\mu$ in the adjoint
 representation, and averaging over background $B_\mu$ is implied by
 angular brackets (where subscript $B$ will be omitted).

 Referring the reader for details of derivation to  refs. [13,18,21]
 and Appendix 2, one can write for (1) the path integral
 \be
 G_{\mu\nu,\mu'\nu'}(x,y|x',y')=const \int^\infty_0 ds \int^\infty_0
 ds' DzDz'e^{-K-K'}\langle\Gamma^{(in)} W_F\Gamma^{(out)}\rangle
 \ee
 where $K=\frac14\int^s_0(\frac{dz}{d\tau})^2 d\tau,~~K'$ is the same
 with primed $z,\tau,s$, and
 \be
 \langle W_F\rangle =tr P_BP_F\langle \exp \{ig \int_C B_\mu du_\mu +2ig
 \int^s_0\hat F d\tau +2ig \int^{s'}_0\hat Fd\tau'\}\rangle
 \ee
  Here
 $P_B,P_F$ are ordering operators of the color matrices $B_\mu$ and
 $\hat F$ respectively. As we shall see in section 3, the terms with
 $\hat F$ generate spin--dependent contribution of nonperturbative
 background, which is calculable and small, and we shall treat those
 terms perturbatively.

 Neglecting $\hat F's $ as a first approximation and
 omitting for simplicity projection operators
 $\Gamma^{(in)},\Gamma^{(out)}$ which do not influence the form of
 resulting Hamiltonian, one arrives at the Wilson loop in the adjoint
 representation, for which one can use the minimal area law,
 confirmed by numerous lattice data [22]  at least up to the distance
 of the order of $1 fm$, \be \langle W_{adj}\rangle =Z~ exp
 (-\sigma_{adj}S_{min}) \ee where we have included in $Z$ self-energy
 and nonasymptotic corrections, since (5) is valid for large loops
 with size $R, T\gg T_g$, where $T_g$ is the gluon correlation
 length.

 Note, that we could treat (4) by the field correlator method
 [13,23], keeping only lowest (Gaussian) correlator $\langle
 F(x)F(y)\rangle $. In this case the leading term will again have the
 form (5); we shall use this method to evaluate contribution of the
 gluon spin terms $\hat F$ in (4); the result (5) is more general,
 since $\sigma_{adj}$ in (5) contains contribution of all correlators
 and is not connected to the Gaussian approximation.

 Applying now the general method of [14] to the Green's function (3),
 one introduces auxiliary (einbein) function $\mu(t)$ of the real time
 $t$ instead of the proper time $s,\tau$, via relation $d\tau
 =\frac{dt}{2\mu(t)}$, and einbein auxiliarly function $\nu(\beta,
 t)$ to  get rid of the square root  Nambu-Goto form for $S_{min}$ in
 (5).
 As a result one    defines the hamiltonian $H$, through the equality
 $G\sim exp (-TH)$, where $T$ is the evolution parameter, taken here
 to be the center-of-mass time $T, 0\leq t\leq T$.

The resulting relativistic Hamiltonian
for two spinless  gluons looks like [14]
$$
H_0=\frac{p_r^2}{\mu(t)}+
\mu(t)+\frac{L(L+1)}{r^2[\mu+2\int^1_0(\beta-\frac12
)^2\nu d\beta]}+
$$
\be
+\int^1_0\frac{\sigma^2_{adj}d\beta}{2\nu(\beta,t)} r^2 +\frac12
\int^1_0 \nu(\beta, t) d\beta
\ee
Here $\mu(t)$ and $\nu(\beta,t)$ are positive auxiliary
functions which are to be found from the  extremum condition [14].
Their extremal values are equal to the  effective gluon energy
$\langle \mu\rangle $ and energy density of the adjoint string $\langle \nu\rangle $.

For the case  $L=0$ the extremization over $\mu$ and $\nu$ yields a
simple answer [14], coinciding with the Hamiltonian of the
relativistic potential model \be H_0=2\sqrt{p^2_r} + \sigma_{adj} r
\ee
An approximation made in  [16-18] corresponds to the replacement of
the operators $\mu(\tau), \nu(\tau,\beta )$ (which by extremization
are expressed through operators $p, r$) by $c$ --numbers, to be found
from extremization of eigenvalues of $H_0$. This yields another form,
used in [18],
\be
H'_0= \frac{\vep^2}{\mu_0} +\mu_0+\sigma _{adj}r
\ee
as can be seen from Table 3 of ref. [19], eigenvalues of (8) are about
5\% higher than those of $H_0$.
The value of $\sigma_{adj}$ in (8) can be found from the string
tension of $q\bar q$ system, since the Casimir scaling found on the
lattice [22] predicts that
\be
\sigma_{adj}= \frac{C_2(adj)}{C_2(fund)} \sigma_{fund} =\frac94
\sigma_{fund }
\ee
For light quarks the value of $\sigma_{fund}$ is found from the slope
of meson Regge trajectories and is equal to
\be \sigma_{fund} =\frac{1}{2\pi\alpha'}\approx 0.18 GeV^2
\ee
>From that we find
\be
\sigma_{adj}\approx 0.40 GeV^2
\ee

In what follows the parameter $\mu$ and its optimal value $\mu_0$, which
enters in (8) play very important role. The way they enter spin
corrections in Section 3 and magnetic moments shows that $\mu_0$
plays the role of effective (constituent) gluon mass (or constituent
quark mass in the equation for the $q\bar q$ system).

In contrast to the potential models, where the constituent mass of gluons and
quarks is introduced as the fixed input parameter in addition to the
parameters of the potential, in our approach $\mu_0$ is calculated from the
extremum of the eigenvalue of equation (8), which yields
$$
\mu_0(n)=\sqrt{\sigma}(\frac{a(n)}{3})^{3/4},
M_0(n)=4\mu_0(n)
$$
where $\sigma=\sigma_{adj}$ for gluons and $\sigma=\sigma_f$ for
massless quarks, and a(n) is the eigenvalue of the reduced equation
$\frac{d^2\psi}{d\rho^2}+(a(n)-\rho-L(L+1)/\rho^2)\psi=0$.
The first several values of $a(n)$ and $\mu_0(n)$ are given in the
Table 1, and will be used in Section 3.

Note that our lowest "constituent gluon mass" $\mu_0(n=L=0)=0.528
GeV$ (for $\sigma_f=0.18 GeV)$ is not far from the values
introduced in the potential models, the drastic difference is that
$\mu_0$ depends on $n,L$ and grows for higher states, and is
calculable in our case.

The eigenvalues of $H_0$, Eq. (7), for $L=0$ and different
$n=0,1,2,...$ are given in Table 2 for $\sigma_{adj}=
\frac94\sigma_{fund}=\frac94 0.18 GeV^2$.
The mass spectrum for $L> 0$ is given by eigenvalues  of
$H_0$, Eq.(6), and was studied in [14]. With the 5\% accuracy of WKB
approximation one can exploit much simpler expressions found in [19],
which predict for $L> 0$ the eigenvalues shown in Table 3.
An independent numerical estimation of the rotating
string spectrum was done in [24] and yields similar eigenvalues.

>From Tables 1,2,3 and [24] one can see that mass spectra of the
Hamiltonian (6) are described with a good accuracy by a very simple
formula
\be
\frac{M^2}{2\pi\sigma}=L+2n_r+c_1
\ee
where $L$ is the orbital momentum, $n_r$ --radial quantum number
and $c_1$ is a constant $\approx 1.55$. It describes an infinite set
of linear Regge-trajectories shifted by $2n_r$ from the leading one
($n_r=0)$.
The only difference between light quarks and gluons at this stage is
the value of $\sigma$, which determines the mass scale.

Thus the lowest glueball state with $L=0, n_r=0$ according to
Table 2 and eq.
(12) has $M^2=4.04 GeV^2.$

It corresponds to a degenerate $0^{++}$ and $2^{++}$ state.
\be
M=2.01 GeV
\ee

In order to compare our results with the corresponding lattice calculations
[1-4 ]
 it is convenient to consider the quantity $\bar M/\sqrt{\sigma_f}$, which
is not sensitive to the choice of string tension $\sigma$
~\footnote[2]{Note that the value $\sigma_f\simeq 0.23 GeV^2$ used in
lattice calculations differs by about 20\% from the "experimental"
value (10).}.

>From these data we have for $L=0, n_r=0$ states the spin averaged mass
\be
\frac{\bar
M}{\sqrt{\sigma_f}}=\frac{M(0^{++})+2M(2^{++})}{3}\frac{1}{\sqrt{\sigma_f}}
\ee
the value $4.61\pm 0.1$, which should be compared to our prediction
$\bar M^{theor}(L=0, n_r=0)/\sqrt{\sigma_f}=4.68.$

For radially excited state our theory predicts
\be
\frac{\bar M^{theor}}{\sqrt{\sigma_f}}(L=0,n_r=1)=7.0
\ee

Lattice data [1] give for this quantity
\be
\frac{\bar M^{lat}}{\sqrt{\sigma_f}}(L=0,n_r=1)=6.56\pm 0.55
\ee

For $L=1, S=1$ states one can define spin-averaged mass in a similar
way
\be
\frac{\bar M}
{\sqrt{\sigma_f}}=\frac{M(0^{-+})+2M(2^{-+})}{3}\frac{1}{\sqrt{\sigma_f}}
\ee
lattice data [1-4] yield
\be
\frac{\bar M^{lat}}{\sqrt{\sigma_f}}(L=1,n_r=0)=6.11\pm 0.38
\ee
which is in a reasonable agreement with our prediction
\be
\frac{\bar M^{theor}}{\sqrt{\sigma_f}}(L=1,n_r=0)=6.0.
\ee

For $L=2,n_r=0$ and $L=1, n_r=1$ we have the spin averaged state  we
have \be \frac{\bar M^{theor}(l=2,n_r=0)}{\sqrt{\sigma_f}}=7.0;
\frac{\bar M^{theor}}{\sqrt{\sigma_f}}(L=1,n_r=1)=8.0
\ee

Lattice data [1] yield respectively $7.7\pm0.4$ and $7.94\pm
0.48$. Note that in the first multiplet
lattice data exist only for $3^{++}$.
 The overall comparison of spin-averaged masses
computed by us and on the lattice is shown in Table 4.

Thus we come to the conclusion that the spin--averaged masses
obtained from purely confining force with relativistic kinematics for
valence gluons are in a good correspondence with lattice data, which
implies that PGE shifts of glueball masses in lattice
calculations is small.

    \section{Spin splittings of glueball masses}

   Here we shall treat spin effects in a perturbative  way; a
   glance at our predictions in Table 5 and at  the lattice results
   given in Table 6 tells that spin splittings in glueball states
   apart from $2^{++}-0^{++}$ amount to less than 10-15\% of the
   total mass, and hence perturbative treatment is justified to this
   accuracy level.

   Before starting actual calculations of spin splittings one should
   choose between two possible strategies (and corresponding physical
   mechanisms) of treating gluon polarizations. In the first approach
   one insists on the            transversality condition and on the
   resulting two gluon polarizations as for the free gluon [7], (this
   is the procedure accepted e.g. in [28]).

   In the second approach it is assumed that gluon has acquired
   nonzero mass due to the adjacent string, similarly to the case of
   $W^\pm, Z^0$, where mass is created by the Higgs condensate. In
   this case one has 3 massive gluon polarizations and the spin
   coupling scheme of two gluons can be taken as the $LS$ scheme with
   the characteristic $J^{PC}$ pattern of lowest levels, which is
   observed in lattice calculations [1-4].

Therefore we choose the second approach and consider gluon spin
   operator $\veS^{(i)}, i=1,2$, total spin operator
   $\veS=\veS^{(1)}+\veS^{(2)}$ and orbital angular momentum $\veL$,
   the total angular momentum ${\bf J}={\bf L}+{\bf S}$, and assign
   to each level (mass) not only conserved values of $J^{PC}$, but
   also values of $L,S$ (which in some cases may have admixture of
   $L'=L\pm 2$, $S'=S\pm 2$, but this admixture in  general case is
   small).

   The detailed discussion of the gluon mass generation in the
   context of gauge invariance and symmetry breaking (as also in the
   electroweak case) is relegated to a separate publication.

   The two--gluon mass operator can be written as
   \be
   M=M_0(n,L) +\veS\veL M_{SL}+\veS^{(1)}\veS^{(2)} M_{SS}+M_T,
   \label{7.1}
   \ee
   where $M_0$ is the eigenvalue of the Hamiltonian $H\equiv
   H_0+\Delta H_{pert}$, and $H_0$ is given in (7) (or its
   approximation in (8)), while $\Delta H_{pert}$ is due to
   perturbative gluon exchanges and discussed in the next section.

   To obtain three other terms in (21) one should consider averaging
   of the operators $\hat F$ in the exponent of (4) and take into
   account that
   \be
   -2i\hat F_{\mu\nu}=2(\veS^{(1)} \veB^{(1)}+ \tilde{\veS}^{(1)}
   \veE^{(1)})_{\mu\nu}
   \label{7.2}
   \ee
   and similarly for the term in the integral $\int\hat F d\tau'$,
   with the replacement of indices $1\to 2$.
Here gluon spin operators are introduced, e.g.
\be
(S^{(1)}_m)_{ik}=-ie_{mik},~~i,k =1,2,3, (\tilde S^{(1)}_m)_{i4}
=-i\delta_{im}
\label{7.3}
\ee

   Two remarks are in order here: {\it i)} the gluon spin enters via
the integral
 $\int 2{\bf S B} d\tau'=\int\frac{{\bf SB}}{\mu(t)} dt $, where
 $\mu(t)$ with its extremum value $\mu_0$ is the same as in (6), (8)
 (for details see Appendix 2).
 {\it ii)} the main part of the Hamiltonian, $H_0$ is diagonal in
 spin indices $i,k$; while the spin--dependent part (22) is treated
 as a perturbation, hence the admixture of the 4-th polarization due
 to $\tilde S$ in (23) does not appear to the lowest order.

 The detailed derivation of spin--dependent terms is done
in the Appendix 3, here we only quote the  results. Since the
structure of the term $\hat F$ in (4) due to (22) is the same as in
case of heavy quark with the replacement of the heavy quark mass  by
the effective gluon parameter $\mu_0$ (see (8)), one can use the spin
analysis of a heavy quarkonia done in [25], to represent the
spin--dependent part of the Hamiltonian in the form similar to that
of Eichten and Feinberg [26]
 $$ \Delta H_s= \frac{\veS\veL}{\mu^2_0}
 (\frac1r \frac{dV_1}{dr} +\frac1r\frac{dV_2}{dr})+
   \frac{\veS^{(1)}\veS^{(2)}}{3\mu^2_0} V_4 (r)+ $$ \be +
   \frac{1}{3\mu_0^2} (3(\veS^{(1)}\ven)(\veS^{(2)}\ven)
   -\veS^{(1)}\veS^{(2)})V_3(r)+\Delta V
   \label{7.4}
   \ee
   where $\Delta V$ contains higher
   cumulant contributions which can be estimated to be   of the
   order of 10\% of the main term in (24) and will be neglected in
   what follows.
     Note that spin of gluon is twice that of quark,
    therefore spin--orbit and spin--spin terms for glueballs are
    effectively twice and four times larger respectively than for the
    quarkonia case.

   The functions $V_i(r)$ are the same as   for heavy quarkonia [25]
   except that  Casimir operators make them $9/4$ times larger;
    the corresponding expressions of $V_i(r)$ in terms of
    correlators $D(x), D_1(x)$ [23], are given in the Appendix 3.
   Both $D$ and $D_1$ are measured on the lattice [27] and $D_1$ is
   found to be much smaller  than $D$. Therefore one can neglect the
   nonperturbative part of $V_3(r)$, while that of $V_4$ turns out to
   be also  small numerically, $M_{SS}(nonpert.) < 30 MeV$, and we
   shall also neglect it.

   The only sizable spin--dependent nonperturbative contribution
   comes from the term $\frac{dV_1}{dr}$ (Thomas precession) and can
   be written at large distances  as
   \be
   \Delta H(Thomas) =
   -\frac{\sigma_{adj}}{r}\frac{\veL\veS}{2\mu^2_0}
   \label{7.5}
   \ee

   Now we come to the point of perturbative contributions to spin
   splittings. The simplest way to calculate those to the order
   $O(\alpha_s)$ (and this procedure holds true  for
   quarkonia) is to represent perturbative gluon exchanges
    by the same Eichten--Feinberg formulas (24) where one
   should keep in $V_i(r)$ only perturbative contributions to
   correlators $D$ and $D_1$ in (A3.8)-(A3.11); then to the order
   $O(\alpha_s)$ one obtains \be \frac1r
   \frac{dV_1^{(pert)}}{dr}=0,~~
    \frac{dV_2^{(pert)}}{dr}=\frac{C_2(adj) \alpha_s}{r^2},
    \label{7.6}
    \ee
    \be
    V_3^{(pert)}=\frac{3C_2(adj) \alpha_s}{r^3},
    \label{7.7}
    \ee
    \be
    V_4^{(pert)}= 8 \pi C_2(adj) \alpha_s\delta^{(3)}(r)
         \label{7.8}
      \ee

However this procedure  should be corrected for glueballs
since $i)$ valence and exchanged gluons are identical and $ii)$ there
is a 4--gluon vertex in addition. The corresponding calculations have
been done in [28], which show that  corrections amount to the
multiplication in (26) with the factor $3/4$ and in (28) with the
factor $5/8$.

With the account of these corrections the
 corresponding matrix elements in (\ref{7.1}) look like
\be
M_{SL}^{(pert)}=
\frac{3C_2(adj)}{4\mu_0^2}\langle \frac{\alpha_s}{r^3}\rangle
\label{7.9}
\ee
\be
M_{SS}^{(pert)}=
\frac{5\pi C_2(adj)}{3\mu_0^2}\langle \alpha_s\delta^{(3)}(r)\rangle
\label{7.10}
\ee
\be
M_{T}^{(pert)}=\frac{C_2(adj)}{\mu_0^2}\langle \frac{\alpha_s}{r^3}
(3\veS^{(1)}\ven\veS^{(2)}\ven -\veS^{(1)}
\veS^{(2)})\rangle
 \label{7.11}
  \ee

>From (\ref{7.10}) one can see that $M_{SS}$ can be written as
\be
M_{SS}=\frac{5\alpha_s}{4\mu^2_0}|R(0)|^2
\label{7.12}
\ee

To make simple estimates, we shall neglect first the interaction due to
PGE between valent gluons. Indeed we show in the
next section that this interaction cannot be written as Coulomb potential
between adjoint charges, and comparison to perturbative
BFKL Pomeron
theory [12] shows that it is much weaker than Coulomb potential.
Neglecting this interaction altogether, one gets the lower bound of
spin-dependent effects, since all matrix clements, like
$\langle\delta^{(3)}
(r)\rangle,\langle\frac{1}{r}\rangle,\langle\frac{1}{r^3}\rangle$
are enhanced by attractive Coulomb interaction.

For purely linear potential
 one has simple relation, not depending on radial
quantum number $n$ [29]
\be
|\Psi(0)|^2=\frac{|R(0)|^2}{4\pi}=\frac{\mu_0\langle
V'(r)\rangle }{4\pi}=\frac{\mu_0\sigma_{adj}}{4\pi}
\ee

Using (33) and $ M_0=4\mu_0$, and taking $M_0$ from Table 1, one
obtains
\be M_{SS}=\frac{5\alpha_s\sigma_{adj}}{M_0},
 \ee
  and for
$n_r=0,1$ and $\alpha_s=0.3$
 the spin-spin splitting is
 \be M_{SS}(n_r=0)=0.3 GeV,
M_{ss}(n_r=1)=0.20 GeV \ee For $M(0^{++})$ and $M(2^{++})$ one has
the values given in Table 5 for $\sigma_f=0.18 GeV^2$ and for the
sake of comparison with lattice calculations in Table 6 for
$\sigma_f=0.228 GeV^2$ and $\alpha_s=0.3 $.

For $L > 0$ one needs to compute spin corrections $M_{SL}$ and
$M_T$. First of all one can simplify matter using the equation (it is
 derived in the same way, as (33) was derived in [29], for details
see Appendix 4))
 \be
L(L+1)\langle\frac{1}{r^3}\rangle=\frac{\mu_0}{2}\langle V'(r)\rangle
\ee
For $V(r)=\sigma_{adj}r$ both $M_{SL}^{(pert)}$ and $M_T^{(pert)}$
are easily calculated and listed in Table 7.

The nonperturbative part of spin   splittings is due to the Thomas
term, ($\frac{d V_1}{dr}+\frac{dV_2}{dr}$), and is calculated
numerically using the exponential form of $D, D_1$ found on the
lattice [27], for details see [25].

The resulting figures for $\triangle M_{Thomas}$ are given in Table
7.  Combining all corrections and values of $M_0$ from Table 2,3 one
obtains glueball masses shown in Table 5 for $\sigma_f=0.18 GeV^2$
and compared with lattice data in Table 6 for $\sigma_f=0.228 GeV^2$.

One can see in Table 6 that calculated spin splittings of lowest
levels are in good agreement with lattice data. This is another
phenomenological manifestation of the PGE suppression in the glueball
system; indeed had we taken PGE in the adjoint Coulomb form with
$\alpha_s=0.3$, we would obtain 3 times larger spin splittings [18].

   The general feature of spin--dependent contribution $\Delta H_s$
   is that it dies out very fast with the growing orbital or radial
   number, which can be seen in the appearance of the $\mu^2_0$
   factor in the denominator of (29-31).

   Indeed, from (8) one can derive that  $M_0\approx 4\mu_0$ and
   therefore   $\Delta H_s\sim
    \frac{1}{M^2(n,L)}\langle O(\frac1r)\rangle $, where
   $O$ stands for terms like $const. \frac1r$ or $const'. \frac{1}
   {r^3}$
   (from perturbation theory). Hence spin splittings of the radial
   recurrence of states $0^{++}, 2^{++}$ or $0^{-+}, 2^{-+}$ should
   be  smaller than the corresponding ground states.
   This feature is also well supported by the lattice data in Table 5.

\section {Perturbative gluon ladders and glueballs}

In many analytic calculations of glueball masses it is postulated that there
is a Coulomb-type interaction between valence gluons, which differs from the
$q\bar q$ case by the Casimir factor, $C_2(adj)=3$ instead of
$C_2(fund)=\frac{4}{3}$. Before going into the details of the question how
the perturbative gluon exchanges give rise to the Coulomb kernel, we
here first assume that this is indeed the case, and correspondingly
calculate the eigenvalues of the hamiltonian
 \be
H=H'_0-\frac{C_2(adj)\alpha_s}{r}
\ee
where $H'_0$ is given in (8). The resulting masses are listed in Table
8 (the
first three lines) for $\alpha_s=0, 0.2, 0.3, 0.39$.

One can see a drastic decrease of the mass due to the Coulomb attraction,
especially for L=0. For a conservative value $\alpha_s=0.3$ this mass drops
down by 0.5 GeV.

This is much larger than in the $q\bar q$ case [13,19], evidently
due to the large Casimir factor.

Another characteristics of the Coulomb shift, which is useful for the
comparison with perturbative Pomeron approach [12], is the Regge
slope $\alpha'_G(0)$ and Regge intercept $\alpha_G(0)$ of the
glueball trajectory drawn as a straight line through the glueballs
with $L=0 (2^{++})$ and $L=0 (4^{++})$ \footnote{This discussion is
rather qualitative. Indeed Coulomb interaction modifies linearity
of nonperturbative glueball trajectories.}. These values are given in
the last two lines of Table 8, and show a drastic increase of the
intercept due to Coulomb interaction by $\Delta\alpha_G(0)\approx
0.64$ for $\alpha_s=0.3$.

This will be compared later in this chapter with a similar large
shift of the perturbative Pomeron trajectory $\Delta\alpha_P(0)$
in the lowest $O(\alpha_s)$ approximation [12] and with much smaller
value of $\Delta\alpha_P(0)$ in the next (one--loop)
approximation [30].

This comparison casts
one more doubt on the validity of the assumption about
the presence of the adjoint Coulomb interaction
 in the form (37).

 A similar conclusion can be deduced from spin-averaged
eigenvalues.  Indeed for the Hamiltonian
\be
 H=H'_0-\frac{C_2(adj)
\alpha_s}{r} \label{7.15}
\ee the eigenvalues are given in Table 8.

One can see that for $L=0$ both $\alpha_s=0.3$ and $0.39$ strongly
contradict data, which shows that perturbative gluon ladder strongly
differs from the adjoint Coulomb interaction,  moreover the
overall agreement of our results for $M_0$ (where no Coulomb
interaction is present)
 with spin averaged lattice masses tells that PGE
 is strongly reduced on the lattice.

 To study this point in detail   one should consider the
 set of perturbative gluon exchanges and compare them to the BFKL
  diagrams describing the perturbative Pomeron [12].

First of all one should look into the mechanism which produces color Coulomb
interaction, and it is instructive to compare quark-antiquark and gluon -
gluon system from this point of view. For both systems there are diagrams of
gluon exchanges in the order $O(g^2)$, and in addition for gluon -
gluon system there is in the same order the diagram of contact
interaction, which affects the hyperfine splitting [28].

The main point is whether and how these diagrams are summed up to produce the
color Coulomb kernel in the exponent, entering the Green's function
of the system. For the $q\bar q$ system (neglecting spin degrees of
freedom for simplicity of comparison) one has the exact
Feynman-Schwinger representation
\be
 G_{q\bar q}=\int dsd\bar s
DzD{\bar z}e^{-K-\bar K}\langle W(C_{z\bar z})\rangle \ee where the
Wilson loop is along the paths $z,\bar z$ integrated in (39). One
can use the cluster expansion for purely perturbative gluons in
$W(C)$ as was done e.g. in [13,21]
 \be \langle
W(C)\rangle=
exp[-C_2\frac{g^2}{2}\int\limits_c\int\limits_c\frac{dz_\mu d\bar
z_\mu}{(z-\bar z)^2}+0(g^4)]
\ee

For straight-line trajectories $z(\tau),\bar z(\tau')$ (e.g. for static
quarks) the integral in the exponent
 of (41) readily yields the color Coulomb
potential, $\langle W\rangle \sim exp(\frac{C_2\alpha_s}{r}t)$.

For light quarks one can consider the integral in the exponent of
(40) as the full-fledged relativistic Coulomb kernel. It is
legitimate to keep this kernel which is $O(g^2)$ in the
exponent,neglecting $O(g^4)$ additional terms provided the Coulomb
kernel yields some amplification.

This is indeed true in the nonrelativistic region (where Coulomb correction
are of the order $\frac{\alpha_s}{u},u\ll 1$) or at small distances (high
energies) where this kernel yields double logarithmic terms [31].
Let  us turn now to the $gg$ system (the same is true a fortiori for
the three-gluon system).

In (3) we have derived the $gg$ Green's function for valent
perturbative gluons in the nonperturbative background. The similarity
of forms (3) and (39) is only superficial, since the Wilson loop in
(39) contains both perturbative  and nonperturbative contributions
and one may argue, that perturbative exchanges dominate at small
distances, and hence exponentiate as in (40) and consequently give
forth the color Coulomb kernel.

In contrast to that , in (3) $\langle W_F\rangle$ contains only
nonperturbative  fields $B_{\mu}$, yielding confining string between
 gluons, but no perturbative exchanges at all. In the  framework of
 the background perturbation theory the perturbative vertices
 $0(a^3)$ and $0(a^4)$ enter the interaction Lagrangian, and there is
 a priori no guarantee that gluon exchanges produced by these
 vertices exponentiate to give a color Coulomb kernel. (Note that
 there is a difference between gluon exchanges, and spin - dependent
 vertices considered in the previous chapter, since the latter are
taken as a perturbation in the lowest order, and there is no need
 for them to exponentiate to the Coulomb ladder).

Having all this in mind, we turn our attention to the subset of graphs which
 is summed up in the BFKL approach [12], and which is distinguished
 by the principle of leading diagrams in the high--energy scattering,
 or in another setting, by the summation of  ladders for the leading
 Regge trajectory  in the $t$--channel. Since these ladders are
  dominant perturbative series (see [12]) for the Pomeron
 trajectory, we can consider the same contribution in our
 circumstances - for calculation of glueball masses, extending in
 this way BFKL -- type analysis from Pomeron--generating glueballs
 ($4^{++},2^{++} etc$) to all others, and having in mind, that this
 may give only an estimate of the order of magnitude.

 Thus our purpose is now to estimate the contribution of the BFKL diagrams to
 the glueball masses (perturbative mass shift) and compare it to the usual
 color Coulomb contribution.

 In order to estimate effects of small distance contributions we shall use
 the analysis of these effects on gluonic Regge--trajectories not from the
 glueballs mass spectra at positive $t$, but for $t=0$. Extensive
 calculations of the gluonic Pomeron trajectory intercept have been
 carried out in the leading log approximation (LLA) [12]  and
$\alpha_s$ corrections were calculated recently [30]. It has been
shown that the leading Regge singularity corresponds to a sum of
ladder type diagrams, where exchanged gluons are
 reggeized. In the leading approximation an intercept of this
 singularity is equal to [12]
 \be \alpha_P (0)=1+\alpha_s\frac{4
 N_c}{\pi}ln 2
 \ee
  The shift from the noninteracting gluons point
 $\alpha_P(0)=1$ is equal to   $\Delta=\alpha_P(0)-1\approx
 0.5$ for $\alpha_s\approx 0.2$. This rather large shift is strongly
 reduced by $\alpha_s$ corrections
 [30]
  \be
 \Delta=\alpha_s\frac{12}{\pi}ln2(1-C\alpha_s)
 \ee

The coefficient $C$ is rather large $(\approx 6.5)$ and the
$\alpha_s$ correction strongly reduces $\Delta$. Its value depends on
 the renormalization scheme and scale for $\alpha_s$. In the
"physical" (BLM) scheme values of $\Delta$ are in the region
 $0.15\div 0.17$ [30]. In this approximation the leading gluonic
 singularity is Regge-pole. and we can estimate mass-shift of the
 lowest glueball state using this result and assuming that the slope
 $\alpha'_P=
\frac{1}{2\pi\sigma_{adj}}\approx 0.4 GeV^{-2} (11)$
will not be strongly modified by perturbative effects. Thus one can
expect that characteristic shift due to perturbative effects in $\bar
M^2(L=0,n_r=0),\delta\bar M^2\approx
\Delta/\alpha'_{adj}\approx(0.38\div 0.48)GeV^2$. This corresponds to the
shift in $\bar M(L=0,n_r=0),\delta\bar M\approx \delta\bar M^2/2\bar
M\approx 0.1 GeV$. This shift should be compared with much larger
mass shift from pure Coulomb interaction given in Table 8.
Thus the $O(\alpha_s)$ correction to the BFKL ladder gives a strong
suppression of PGE series and may be a possible explanation why
Coulomb --like attraction is not seen neither in spin--averaged
masses $\bar M(L, n_r)$, nor in spin splittings. It should be noted
that this is only a rough estimate of the perturbative effects
because higher orders of the perturbation theory can modify this
result.

  \section{Three--gluon glueballs}

 The three--gluon system can be considered in the same way, as it
 was done for the two-gluon glueballs. The $3g$ Green's function
 $G^{(3g)}$ is obtained as the background--averaged product of 3
 one--gluon Green's function, in full analogy with (1). Assuming large
 $N_c$ limit for simplicity and neglecting spin splittings
 and projection operators one
 arrives at the path integral  (cf equation (3))
 \be
 G^{(3g)}=const \prod^3_{i=1}\int^\infty_0 ds_i  Dz^{(i)}
 e^{-K_i-\sigma S_i}
 \ee
 where $\sigma \equiv \sigma_{fund}
 $, since every gluon is connected by a fundamental string with each
 of his neighbors.

 Using as before  the  method of
 [13,14]
 and three--body treatment of [17] one obtains  omitting
 spin--dependent terms the following Hamiltonian (we assume symmetric
 solution with equal $\mu_i(\tau)\equiv \mu(\tau), i=1,2,3$ (no
 orbital excitations was assumed as in (8)) \be
 H^{(3g)}=\frac{\vep^2_\eta+\vep^2_\xi}{2\mu}+\frac{3\mu}{2}+
 \sigma\sum^3_{i< j=1}r_{ij}
 \ee
 Here $r_{ij}=|\ver_i-\ver_j|$, and $\ver_i$ is the space coordinate
 of the i-th gluon, while $\vexi,\vep_xi$ and $\veet,\vep_\eta $ are
 defined as
 \be
 \veet=\frac{\ver_1-\ver_2}{\sqrt{2}},~~
 \vexi=\sqrt{\frac{3}{2}}
 (\frac{\ver_1+\ver_2}{2}-\ver_3), ~~
 \vep_\xi=\frac{1}{i}\frac{\partial}{\partial_{\vexi}},~~
 \vep_\eta=\frac{1}{i}\frac{\partial}{\partial_{\veet}}
 \ee

To simplify treatment further, we shall consider $\mu$ as a constant to
be found from the extremum of eigenvalues, as in (8), which in that
case provided some 5\% increase in eigenvalues (see Table 3 of
[19]), and what we expect also in this case.

To find eigenvalues of $H^{(3g)}$ one can use the hyperspherical
method introduced in [32] and applied to the $3q$ system in [17].
Defining hyperradius $\rho $, $\rho^2= \veet^2+\vexi^2$,
one obtains one--dimensional equation for the eigenfunction
$\chi^K_n(\rho)\equiv \chi(\rho)$,($K$ is the grand angular momentum
 $K=0,1,2...$,
and $n$ --radial quantum number).  \be
-\frac{1}{2\mu}\chi^{\prime\prime}+U_{eff}(\rho) \chi(\rho)
=M\chi(\rho)
\ee
where
\be
U_{eff}(\rho)
=\frac{1}{2\mu\rho^2}(K^2+4K+\frac{15}{4})+\frac{32\sqrt{2}}{5\pi}
\rho\sigma
\ee
Solution of (46) is expressed through generalized Airy functions.

A reliable (within few percent of accuracy) estimate of $M$ is
obtained when  one replaces  $U_{eff}(\rho)$ by the oscillator well,
 with  the center at the minimum of
$U_{eff}(\rho),~~ \rho=\rho_0$, and frequency $\omega_0$
expressed through $U^{\prime\prime}_{eff}(\rho_0)$. For $K=0$
one has
\be
\rho_0=(\frac{75\pi}{128\sqrt{2}\mu\sigma})^{1/3},~~
\omega_0=(\frac{45}{4\mu^2\rho^4_0})^{1/2}
\ee
In this way one obtains
\be
M(\mu) =\frac{3\mu}{2} + 36( \frac{\sigma^2}{45 \mu\pi^2})^{1/3}+
\frac{\omega_0}{2}
\ee
and minimization of $M(\mu)$ in $\mu$ yields
\be
\mu_0\cong (1.6)^{1/4}(\frac{8\sigma}{3\pi})^{1/2};~~
M(\mu_0)=6\mu_0+\frac{\omega_0}{2}
=6\mu_0+(\frac85)^{3/4}(\frac{3\sigma}{\pi})^{1/2}
\ee

For $\sigma=0.18 GeV^2 $ one obtains $\omega_0=1.18 GeV$ and
    $\mu_0=0.44 GeV$, hence the minimal  eigenvalue is
    \be
    M_0=3.23 GeV
    \ee
    This spin--averaged value is listed in Table 4.
    Radial excitations are given by an approximate equation
    \be
    M_n= 6\mu_0+\frac{\omega_0}{2} +n\omega_0,~~ n=0,1,2...
    \ee
    Orbital excitations yield increase in mass of the order
    \be
    \Delta M^K\sim \frac{K^2+4K}{2\mu\rho^2_0},
    \ee
    which for lowest excitation gives
    \be
    \Delta M^{K=1}\sim 1 GeV
    \ee
    almost the same amount, as for the radial excitation.

    Coulomb shift (if Coulomb interaction existed between gluons),
    would be enormous: $\Delta M(Coulomb) =-1.3 GeV$. However here
    one can use the same arguments as for two--gluon glueballs and
    drop the color Coulomb interaction between gluons altogether.

    Finally we turn to  the question of quantum numbers and spin
     splittings of the $3g$ states.

      According to (27), perturbative  hyperfine interaction is given
       by matrix elements
        \be
        \Delta M_{SS}=\sum_{i>
j}\langle \veS^{(i)}\veS^{(j)}
        \frac{5\pi C_2 (fund)}{3\mu_0^2}
         \alpha_s\delta^{(3)}(\ver_{ij})\rangle
         \ee

Note that in the large $N_c$ limit  gluon lines are replaced by
double fundamental lines and planar gluon exchanges occur with
fundamental charge, hence the fundamental Casimir operator in (55).

For the $K=0$ state the wave function depends only on hyperradius
$\rho$ and one has
\be
\langle \delta^{(3)} (\ver_{ij})\rangle=\langle
\frac{\sqrt{2}}{\pi^2\rho^3}\rangle
\ee

Now for $K=0$ state all internal angular momenta are zero and one can
express
$\langle \veS_i\veS_j\rangle$
 through total angular momentum
$J$
\be
\langle \veS_i\veS_j\rangle = \frac{J(J+1)-6}{6}
\ee
As a result one obtains $\Delta M_{SS}$ for $\alpha_s=0.3$ and
$\sigma =0.18 GeV^2$
\be
\Delta M_{SS}\cong \frac{5\sqrt{2}C_2(fund) \alpha_s}{\pi
\mu^2_0\rho^3_0}\langle \veS_i\veS_j\rangle \approx 0.644\mu_0
\frac{J(J+1)-6}{6} GeV \ee

Hence the candidate for the odderon state $J^{PC}=3^{--}$ is shifted
by $0.28 GeV$ upwards
and  $1^{--}$ by  $0.189 GeV$ downwards with respect to (51).
Resulting values of glueball masses are listed in Tables 5 and 6.

\section{Glueball Regge trajectories and Pomeron.}

The leading Regge trajectory (with the largest intercept $\alpha_P(0)$) is
usually called the Pomeron trajectory. It plays a special role in the reggeon
approach to high-energy hadronic interactions. The parameters of the Pomeron
trajectory and especially its intercept play a fundamental role for
asymptotic behaviour of diffractive processes. We already touched the problem
of the Pomeron intercept in Section 4, where the perturbative, small
distance contribution has been discussed. Now we will consider this
problem in more details, taking into account both  nonperturbative
and perturbative contributions to the Pomeron dynamics.

The large distance, nonperturbative contribution gives according to Eq.(12)
for the leading glueball trajectory ($n_r=0$)
\be
\alpha_P(t)=-c_1+\alpha'_P t
\ee
with $\alpha'_P=\frac{1}{2\pi\sigma_{adj}}$

Taking into account spins of "constituent" gluons
but neglecting small nonperturbative spin effects we obtain for
the intercept
of the leading trajectory
\be
\alpha_P(0)=-c_1+2
\ee
which leads to $\alpha_P(0)\approx 0.5$, and this value is
substantially below the value found from analysis of high-energy
interactions $\alpha_P(0)=1.1\div 1.2$ [20].

The perturbative (BFKL) contribution leads to a shift (increase) of
the Pomeron intercept by 0.2, as it was explained in Section 4. The
resulting $\alpha^{(0)}_P(0)\approx 0.7$ is still far from
experiment.

There are other nonperturbative
 sources, which can lead to an increase of the
Pomeron intercept.
One of the most important is in our opinion the quark-gluon
mixing or account of quark-loops in the gluon "medium". In the $1/N$
-expansion the effect is proportional to $N_f/N_c$, where $N_f$ is
the number of light flavours and  this mixing is known to be
important (at least in the small-$t$ region).

In the leading approximation of the $1/N_c$-expansion there are 3
Regge-trajectories, -- $q\bar q$-planar trajectories ($f^{(0)}$ made
of $u\bar u$ and $d\bar d$ quarks and $f'^{(0)}$ made of $s\bar
s$-quarks) and pure gluonic trajectory - $G$. The transitions
between quarks and gluons $\sim\frac{1}{N_c}$ will lead to mixing of
all these trajectories.  Lacking calculation of these effects in QCD
we will consider them in a semi--phenomenological manner. From the
mixing of two trajectories 1 and 2 with the transition constant
$g_{12}$ it is easy to obtain the following values of new
trajectories (see for example [33]) \be
\alpha^{(t)}_{\pm}=\frac{\alpha_1(t)+\alpha_2(t)\pm
\sqrt{(\alpha_1(t)-\alpha_2(t))^2+4g^2_{12}(t)}}{2}
\ee

Note that for a realistic case of $G, f$ and $f'$--trajectories
(Fig.1) all 3--trajectories before mixing are close to each other in
the small $t$ region.  Trajectory of gluonium crosses planar $f$ and
$f'$--trajectories in the positive $t$ region $(t<1 GeV^2)$. In this
region mixing between trajectories is essential even for small
coupling matrix $g_{ik}(t).$

The dual unitarization scheme [34-36] leads to the conclusion
that the quantity $g^2_{12}$ fastly decreases as t increases in the
positive $t$ region. This means that at large positive $t$,
$\alpha_{\pm}(t)$ coincide with trajectories $\alpha_1$ and
$\alpha_2$, as it
happens in (61)  for $g^2_{12}\ll|\alpha_1-\alpha_2|$
 \be
\alpha_+\approx\alpha_1+\frac{g^2_{12}}{\alpha_1-\alpha_2};
\alpha_-\approx\alpha_2+\frac{g^2_{12}}{\alpha_2-\alpha_1}
\ee

This phenomenon is called the asymptotic planarity [36]. Let us note
that mixing effects will be small in the large $t$ region even if the
couplings have weak $t$-dependence because the differences between
planar and gluonic trajectories increase $\sim t$ at large $t$.

For weak mixing between trajectories
$(g^2_{ik}\ll|\alpha_i^{(0)}-\alpha_k^{(0)}|)$ (62) can be generalized
\be
\alpha_i\approx\alpha_i^{(0)}+\sum\limits_k
\frac{g^2_{ik}}{\alpha_i^{(0)}-\alpha_k^{(0)}}
\ee

For $g^2_i\sim 0.1$ typical resulting trajectories are shown in
Fig.1 by solid lines. For details see appendix 5. The Pomeron
trajectory is shifted to the values $\alpha_P(0)\ge 1$. For $t> 1
GeV^2$ Pomeron trajectory is very close to the planar $f$ trajectory.

Position of the second vacuum trajectory at $t\le 0$ is close to
$\alpha_f$, while for $t>1 GeV^2$ it is close to $\alpha_{f'}$.
The third vacuum trajectory is below $\alpha_{f'}$ at $t\le 0$ and at
$t>1 GeV$ it is close to the $\alpha_G$. Due to asymptotic
planarity effects of mixing are not very essential for properties of
physical particles on these trajectories as all resonances are in the
region $t>1.5 GeV^2$. On the other hand they are important for
understanding of the SU(3)-breaking effects for the Pomeron exchange
amplitudes at $t\le 0$.

At the end of this section let us consider the "odderon"--
the leading Regge trajectory with negative signature and $C$-parity.
Mass of the lowest $3g$ glueball with spin $3^{--}$ corresponding to
this trajectory has been estimated in the previous section and found
to be large $\approx 3.51  GeV$ ( for $\sigma_f=0.18 GeV^2$) in
accord with lattice data.  The slope $\alpha'_{3g}$ for this
trajectory should be equal to the one of
$gg$--trajectory\footnote{The situation is analogous here to the case
of $q\bar q$ (meson) and $qqq$ (baryon) Regge trajectories, where
baryon  trajectory displays the quark--diquark structure and hence
the meson Regge slope [17]. } and thus the intercept of the
nonperturbative glueball "odderon" is very low $\alpha_{3g}(0)\approx
-1.5$.  Mixing with $q\bar q$--trajectories $(\omega,\varphi)$ is
much smaller than in the Pomeron case since there is no crossing of
the odderon and $(\omega,\varphi)$ trajectories in the small
$t$-region and thus gluonic "odderon" is not important for high --
energy phenomenology (at least in the small-$t$ region).

\section{Discussion and conclusions.}

The main results of the paper can be separated in two groups. In the
first part we calculate the $2g$ and $3g$ glueball spectrum
analytically and compare resulting masses with lattice data, finding
a very good agreement. In the second part the glueball Regge
trajectories are obtained  and their correspondence with the
Pomeron and odderon is discussed.

Concerning the glueball spectrum the spin-averaged results of section 3
calculated
  for all
states of $2g$ and $3g$ glueballs
 yield a very good agreement
between our results and
spin-averaged lattice masses. We stress that our spectrum in contrast
to most existing theoretical models contains no fitting
parameters, and all masses are expressed in terms of string tension
$\sigma$, as it is done also on the lattice.

This coincidence
 and evident smallness of PGE interaction, which would be
very strong
if it had been adjoint Coulomb interaction due to the Casimir factor
3, has called us for more detailed investigation, whether Coulomb is
indeed appropriate in the systems of valence gluons. The analysis
done in Section 4 has brought us to the conclusion that the
situation in the system of valence gluons is completely different
from that of valence quarks, and the perturbative gluon exchanges do
not exponentiate into Coulomb kernel for $2g$ and $3g$ systems in
contrast to the $q\bar q,3q$ systems.

This observation explains qualitatively the absence of strong Coulomb
downward shifts of glueball masses  and moderate spin splittings in
lattice calculations.  To make a quantitative estimate, we have
considered the BFKL perturbative series for Pomeron [12]
including one loop correction [30].  This series is not a Coulomb
ladder and with an account of NLO corrections it leads to a mass
shift of approximately 3--4 times smaller than for Coulomb
interaction.

In contrast to Coulomb interaction the spin splittings of glueball masses are
obtained from the first perturbative correction calculated with
nonperturbative wave functions. There is a good agreement on spin splittings
(within few tens of  MeV) between our calculations and lattice data,
as it is shown in Table.6.

The agreement signifies that the main ingredient of the glueball dynamics is
the adjoint string (or two fundamental strings) occurring between gluons in
the two-gluon glueballs and the triangle construction of fundamental strings
in the $3g$ glueballs. The string dynamics implies that glueball
masses lie on the corresponding straight-line Regge trajectories,
which have the Regge slope equal to $\frac{4}{9}$ of that for meson
trajectories. In other respects glueball trajectories are similar to
the $q\bar q$ trajectories for zero-mass quarks [13-16,
19,24]:  they are straight with good accuracy, have
daughter recurrences due to radial excitations, which are also
approximately straight lines.

In the last part of the paper we used our knowledge of glueball Regge
trajectories for investigation of the Pomeron singularity. The
Pomeron, which yields asymptotically dominant contribution at large
energies, is certainly a complicated object, which has some features
connecting it with the dominant glueball trajectory.  First of all,
Pomeron exchange has a cylinder topology (which is supported by the
multiplicity analysis [37]), similar to that of glueball amplitude
which becomes evident when one replaces the adjoint string by the
double fundamental string.

            The idea of Pomeron as a two-gluon exchange amplitude
            has a long history [11] and was exploited both in
             perturbative [12], nonperturbative [18] and  hybrid [38]
             approaches. The purely perturbative approach has some
            difficulties of internal consistency both because of slow
            convergence of perturbative series [30] and because of
            sensitivity to large-distance contributions [39]. The
latter signifies that the nonperturbative effects may play very
            important  role in the Pomeron dynamics, and our paper is
            a demonstration of this. The character of nonperturbative
            trajectories is linear due to the string dynamics and
            absence of mass dimension parameter other than string
            tension.  Perturbative singularities in the $j$-plane are
            not always poles, and are certainly not linear
            trajectories.

            Our discussion in previous sections has come to the conclusion
            that perturbative effects only slightly shift nonperturbative
            trajectories (increasing Regge intercept by$\sim 0.2$).

            With all that we have come to the Regge intercept of leading
            Regge trajectory equal to 0.7, strongly different from the
            experimental Pomeron intercept of $1.07\div 1.2$. And
            here comes an interesting observation, made earlier in a
            bit different context [33], that due to different slopes
            of meson and glueball trajectories they must intersect
            each other in the region of small $t<1 GeV^2$. We have
            taken this fact into account in the three-pole model,
            where coupling constants between channels $f,f'$ and $G$
            are introduced phenomenologically.

            The results, shown in Fig.1 demonstrate a dramatic
            change in the course of trajectories: the largest
            intercept increases by 0.5 reaching the physically
            reasonable values of 1.2
             \footnote{Note that we discuss
here the "bare Pomeron" intercept, which is larger than an
"effective" Pomeron intercept usually extracted from HE scattering.
            As was discussed in [20] the bare Pomeron
            characteristics are measured in small-$x$ DIS
            experiments, yielding the intercept around 1.2.}.

            Let us note that both nonperturbative (string dynamics, quark
            loops) and perturbative effects are important to obtain
            $\alpha_P(0)>1$. It is impossible to separate out "soft"
            and "hard" Pomerons, as sometimes is done in
            phenomenological studies of high-energy interactions of
            hadrons and small--x physics of deep inelastic
            scattering.

In a supercritical Pomeron theory with $\Delta\equiv\alpha_P(0)-1>0$
the corresponding multipomeron exchanges are important at very high
energies.  They allow to obtain scattering amplitudes, which satisfy
$s$-channel unitarity and the Froissart bound for total interaction
cross sections as $s\to \infty$.  From the point of view of $1/N_c$
 -- expansion multipomeron exchanges are $\sim (1/N_c^2)^{2n}$ ,
 where n is the number of exchanged Pomerons, but they have faster
 increase with energy $(\sim (\frac{s}{s_0})^{n\Delta})$ than the
pole term and should be resummed.  This can be done using Gribov's
reggeon--diagram technique [40]. In practical applications of
Reggeon theory to a description of high-energy hadronic interactions
multipomeron exchanges are essential for simultaneous description of
total interaction cross sections and multiparticle production (for
review see [41]).

            Looking back to a structure of vacuum trajectories we
            found that each of 3 new trajectories $\alpha_i(t)$ is
            now a mixture of $G,f,f'$ and only asymptotically at
            large $t$ they tend to the original trajectories.  As it
            is seen in Fig.1 the leading trajectory (with largest
            intercept), which should be associated with Pomeron,
            asymptotically tends to $f$, the second trajectory
            as positive $t$  is
             close to $f'$, while the third asymptotically  (at large $t$)
            coincides with $G$ while at $t=0$ it is
            below the first two trajectories.  Thus a rearrangement
            takes place:  $G $
            trajectory is shifted downwards, while the $f$ trajectory
            is lifted up and becomes the Pomeron.

            One of immediate consequences of this rearrangement is the
            special pattern of Pomeron couplings, which can be
            measured experimentally. While $G$ trajectory was flavour
            blind, now due to mixing one can calculate the couplings
            of the Pomeron to light quarks (via $f$), to strange
            quarks (via $f'$) and symmetrically to all flavours (via
            $G$).

            Mixing between gluons and $q\bar q$ pairs has another important
            aspect, -- it leads not only to shifts of $Re
            \alpha_i(t)$ but also to appearance of $Jm \alpha_G(t)$
            and as a consequence to  nonzero   widths of resonances on
            glueball trajectories. They should be of the same size
            as mass shift due to the mixing and thus are expected to
            be not too large, $\Gamma_G\sim 100 MeV$.

            Present study can
            be improved in several points.  Firstly, perturbative
            contributions to the glueball trajectory including
            spin-dependent terms should be studied more systematically.

            Secondly, analytic calculations of $g_{ik}(t)$ are necessary to
            make our theory complete. Finally a detailed analysis of
            experimental consequences of our results is needed, which
            is planned for separate publication.

             The work of A.B.Kaidalov was partly supported  by the
             grants RFBR 98-02-17463 and NATO grant  OUTR.LG 971390.

\newpage

{\Large \bf  Appendix 1} \\

{\bf
Creation operators of glueball states.}      \\

\setcounter{equation}{0} \def\theequation{A1.\arabic{equation}}
 \vspace{1cm}

We consider in this Appendix  and Tables 9,10 the operators
$\Psi^{(in)}_k$ and $\Psi^{(out)}_k$ in (1) and (A2.16)-(A2.17) which
specify glueball states and their quantum numbers, $J^{PC}$. One may
consider local $\Psi(x,x)$ or nonlocal operators $\Psi(x,y)$ for
two--gluon glueballs and corresponding operators for many--gluon
glueballs $\Psi(x^{(1)},...x^{(n)})$.  For simplicity listed below
are only local versions, since nonlocal ones can be  constructed with
the help of parallel transporters $\Phi(x,y)$, as it is done in
(A2.16), (A2.17).

First one can construct $\Psi_k$    in a general form, not assuming
separation of $A_\mu$ into background and valence parts, similarly to
what is done on the lattice. Then one has at his disposal vectors
${\bf E}_a, {\bf D}_a$ pseudovector ${\bf B}_a$ and color tensors
$\delta_{ab}, f_{abc}, d_{abc}$. One should also take into account
that under charge conjugation $C$ the following transformations hold
$$
CA_\mu C^{-1}\equiv A_\mu^{(C)}= -A^T_\mu,
$$
\be
C F_{\mu\nu}C^{-1}=- F_{\mu\nu}^T, ~~ CD_\mu C^{-1}  =-D_\mu^T
\ee

Hence one obtains the following list of states for the two--gluon
glueballs (containing two field operators) and due to Bose statistics
symmetric with respect to exchange of all coordinates of two gluons.
We also list in the first column
of Tables 9,10  the dimension of the corresponding
operator.

In the  background perturbation theory (BPT)   $\Psi^{(in)}$ and
$\Psi^{(out)}$ can be constructed from the spacial components of the
gluonic field $a_i, i=1,2,3$ since the fourth component $a_4$ can be
expressed via the background gauge condition $D_\mu a_\mu=0$. Note
that $a_i$ transforms homogeneously (see equation (A2.4) of Appendix
2) and therefore one obtains gauge--invariant combinations for
$\Psi^{(in)}, \Psi^{(out)}$ replacing $E_i$ in the third column of
the Table 9,10 by $a_i$ whereas $J^{PC}$ does not change. In the same
way $B_k$ is replaced by $({\bf D}\times {\bf a})_k$, and one obtains
the fourth column of the Table 9,10. Dimension of BPT operators is
given in column 5 and the orbital angular momentum in the last
column.

For three--gluon glueballs the corresponding entries are given in
Table 10. One should notice that $C$ parity of all listed states is
here negative. Again, dimension of BPT operators is listed in the
last column.

As can be seen from the results of our calculations in Tables 4-6, the
glueball spectrum is in good agreement with the hierarchy  associated
with increasing angular  momentum $L$ or BPT dimension (they differ
by two units for two--gluon glueballs) The same ordering persists in
lattice data. Three--gluon glueball masses are typically shifted by
$1.5\div 2 GeV$ (an exception of $1^{+-}, 3^{+-}$ states in lattice
data waits for explanation).
Note the absence of $J=1^{++}$ states in the lattice spectrum. From
Table 9 one can see the only candidate, $1^{-+}$, but the
corresponding  local operator is proportional to the energy--momentum
tensor and by the arguments of [42] the residue  of the state should
vanish.

In both Tables 9,10 we have used following notations
$$
symm_{(ik)}T_{ik}= T_{ik}+T_{ki}-\frac{2}{3}\delta_{ik} T_{ll}
$$
Symmetrization of higher operators $T_{ikl}, T_{iklm}$ is done in the
usual way to construct irreducible  $O(3)$ tensors.\\

{\Large \bf  Appendix 2} \\

{\bf Glueball Green's function and Hamiltoman in the background
formalism.}

\setcounter{equation}{0} \def\theequation{A2.\arabic{equation}}
 \vspace{1cm}

In what follows the Euclidean space-time is used.

The total gluonic field $A_\mu$ is split into NP background $B_\mu$ and
valence (perturbative) gluon field $a_\mu$,
\be
A_\mu=B_\mu+a_\mu.
\ee
The QCD partition function $Z(J)$,
\be
Z(J)=\frac{1}{N}\int e^{-S_E(A)+\int J_\mu(x)A_\mu(x)d^4x}DAD\psi
D\bar\psi
\ee
where $S_E$ is can be rewritten using 'tHooft identity as
\be
Z(J)=\frac{1}{N'}\int DB\eta(B)e^{\int{JBd^4x}}\int DaD\psi
D\bar\psi e^{-S_E(B+a)+\int Jad^4x}\ee

Here $\eta(B)$ is (an arbitrary) weight of integration over background fields
$B_\mu$, the exact form of which is not of interest to us, since the overall
effect of background fields will enter in our results via string
tension $\sigma$ and (in some corrections) as a NP field correlator
$\langle F(x)F(y)\rangle$. Both quantities are considered as an
input.

In what follows we shall expand (A2.4) in powers of $ga_\mu$
as it is usually done in
background perturbation theory [21,43]. In the lowest order of
$1/N_c$ expansion quarks are decoupled from gluons and we shall
neglect coupling to quarks till the last two sections of the paper.

It is convenient to prescribe the following gauge transformations,
\be
 a_\mu\to U^+a_\mu U,
 \ee
\be
 B_\mu\to U(B_\mu+\frac{i}{g}\partial_\mu)U,
 \ee
and to impose on $a_\mu$ background gauge condition
\be
D_\mu a_\mu=\partial_\mu a^a_\mu+gf^{abc}B^b_\mu
a^c_\mu=0
\ee

In this case the ghost fields have to be introduced and one can write
the resulting partition function as
\be
 Z(J)=\frac{1}{N'}\int
DB\eta(B)exp(\int{J_\mu B_\mu d^4x})Z(J,B)
\ee
 where
\be
Z(J,B)=\int Da det (\frac{\delta G^a}{\delta\omega^b})exp\int
d^4x[L(a)-\frac{1}{2}(G^a)^2+J^a_\mu a^a_\mu]
\ee
where
$$
L(a)=L_0+L_1(a)+L_2(a)+L_{int}(a)$$
$$
L_2(a)=+\frac{1}{2}a_\nu(\hat D^2_\lambda\delta_{\mu\nu}-\hat D_\mu \hat
D_\nu+ig\hat F_{\mu\nu})a_\mu=
$$
\be
\frac{1}{2}a^c_\nu[
D^{ca}_\lambda D^{ad}_\lambda \delta_{\mu\nu}-D^{ca}_\mu
D^{ad}_\nu-gf^{cad} F^a_{\mu\nu}]a^d_\mu
\ee
$$ D^{ca}_\lambda=
\partial_\lambda\cdot\delta_{ca}+g f^{cda}B^b_\lambda\equiv
\hat D_\lambda$$
$$
L_0=-\frac{1}{4}(F^a_{\mu\nu}(B))^2;~~ L_1=a^c_\nu
D^{ca}_\mu(B)F^a_{\mu\nu}$$
\be
L_{int}=-\frac{1}{2}(D_\mu(B)a_\nu-D_\nu(B)a_\mu)^ag f^{abc}a^b_\mu
a^c_\nu-\frac{1}{4}g^2 f^{abc}a^b_\mu a^c_\nu f^{aef}a^e_\mu
a^f_\nu
\ee

The background gauge condition is written as
\be
G^a=\partial_\mu a^a_\mu+g f^{abc} B^b_\mu a^c_\mu=(D_\mu
a_\mu)^a
\ee
and the ghost vertex [21,43] is obtained from
$\frac{\delta G^a}{\delta\omega^b}(D_\mu(B)D_\mu(B+a))_{ab}$ to be
\be
L_{ghost}=-\Theta^+_a(D_\mu(B)D_\mu(B+a))_{ab}\Theta_b
\ee

The linear part of the Lagrangian $L_1$ disappears if $B_\mu$
satisfies classical equations of motion.

We now can identify the propagator of $a_\mu$ from the quadratic terms in
Lagrangian $L_2(a)-\frac{1}{2\xi}(G^a)^2$
\be
G^{ab}_{\mu\nu}
=[\hat D^2_\lambda\delta_{\mu\nu}-\hat D_\mu \hat D_\nu+ig\hat
F_{\mu\nu}+\frac{1}{\xi}\hat D_\nu \hat D_\mu]^{-1}_{ab}
\ee

It will be convenient sometimes to choose $\xi=1$ and end up with the
well known form of propagator in -- what one would call -- the
background Feynman gauge
\be
G^{ab}_{\mu\nu}=(\hat D^2_\lambda\cdot
\delta_{\mu\nu}+2ig\hat F_{\mu\nu})^{-1}
\ee

We are interested in the glueball Green's function and therefore must define
first the initial and final state vectors of glueballs, consisting of $n_i$
valence gluons initially and of $n_f$ gluons in the final state. The
following general nonlocal state vectors can be used for k gluons
\be
\Psi_k(x^{(0)},
...x^{(k-1)})=
t^r[f_0(a(x^{(0)})\phi(x^{(0)},
x^{(1)})f_1(a(x^{(1)}),.
..f_{k-1}(a(x^{(k-1)})\phi(x^{(k-1)},x^{(0)})]\ee

Here $\phi(x,y)=P~exp~(ig\int\limits^x_y B_\mu(z)dz_\mu)$ is parallel
transporter, all $a_\mu$ are in fundamental representation, and $f(a)$ is a
polynomial in $a_\mu$, which may contain derivatives in the form of
$D_\mu\equiv \partial_\mu-ig B_\mu$.

According to (A2.4),(A2.5) $\Psi_k$ are color singlets. One can also
have local form of $\Psi_k$ taking all $x^{(i)}$ at one point.
Exact form of $\Psi_k$ is given in Appendix 1.

As will be seen below the state (A2.15) will evolve as a closed
fundamental string with k gluons "sitting" on the string, when all
$f_i$ are linear (and more gluons, when some $ f_i$ have larger
power). This form of initial and final states is convenient for
multigluon glueballs and will be used for three-gluon glueballs in
Section 5.

Another form of $\Psi_k$ (equivalent to the previous in the limit
$N_c\to\infty$) obtains when one takes adjoint string. E.g. for two-gluon
glueballs one has
\be
\hat\Psi_2(x^{(1)},x^{(2)})=\hat t_r[\hat a_\mu(x^{(1)}\hat
\phi(x^{(1)},x^{(2)})\hat a_\nu(x^{(2)}]\ee

Here the hat signs imply adjoint representation. For two-gluon glueballs we
are using (A2.17) for initial and final states, and the corresponding
Green's function describes the evolution of the open adjoint string
with adjoint charges (gluons) at the ends.

It can be calculated using (A2.4) in the form given by equation (1)
of the main text, where we neglect terms $L_1(a)$ and $L_{int}(a)$
(the first gives an insignificant correction discussed in [21],
while $L_{int}$ contains higher powers of $ga_\mu$, and will be used
for calculation of perturbative corrections to the Green's function).

The next step is the Feynman--Schwinger representation (FSR)
[13,14] for the gluon Green's function (A2.15), which allows to
exponentiate $B_\mu$ and $\hat F_{\mu\nu}$ as
\be
G_{\mu\nu}(x,y)=const\int\limits^\infty_\infty ds Dz
e^{-K}P_BP_Fe^{(ig\int\limits^x_y\hat B_\mu
dz_\mu+2ig\int\limits^s_0\hat F{(z(\tau))}d\tau)} \ee

Insertion of (A2.17) into (1) and using the fact that
ordering inversion
for one of gluons yields  $(-\hat B^T_\mu,~-\hat F^T)$ instead of
$\hat B_\mu, \hat F$ brings forth the equations (3) and (4) of the
main text.

Equations (3),(4) is another form of (1),(2) and contain no approximations
(except for  omitting of the terms $L_1$ and $L_{int}$
discussed above and used in writing (1) and (2)).

Another important step done first in [13,16], and developed in
[14,15], is the introduction of the auxiliary function
$\mu(\tau)$, which may be called the einbein, and which plays a
crucial role of effective gluon mass in the whole formalism. This is
done rigorously and without introduction of arbitrary fitting
parameters in contrast to usual potential models.

Defining
\be
 2\mu(t)=\frac{dt}{d\tau}, t\equiv z_4
 \ee
where $\tau$ or s is the Schwinger proper time, and t is the Euclidean time
at any point of trajectory $z_\mu(\tau)$, one can rewrite FSR (3)
identically as
\be
 G_{\mu\nu,\mu'\nu'}(x,y|x',y')=const\int
D\mu(t)D\mu(t')Dz_iDz'_k e^{-K-K'}\langle W_F\rangle
\ee
Where $Dz_i$ (or $Dz'_k)$ denote 3d path integrals over trajectories
$z_i(t), i=1,2,3$ and $z'_k(t'), k=1,2,3$, and $D\mu(t)$ is the 1d
path integral over functions $\mu(t)$.  Kinetic terms $K,K'$ can be
expressed through $\mu(t)$, e.g.
\be
 K=\int\limits^T_0
\frac{\mu(t)}{2}[((\dot{z}_i(t))^2+1]dt, T=x_4-y_4
\ee
where dot means time derivative. The form (A2.20) reminds
nonrelativistic kinetic energy however is exact relativistic form.
In case of massive relativistic particle with mass  $m$ in the
corresponding term in action, $K_m$, looks like
\be
K_m=\int\limits^T_0(\frac{m^2}{2\mu(t)}
+\frac{\mu(t)}{2}[(\dot{z}_i(t))^2+1])dt\ee

Introducing momentum $P_i=\frac{\partial K_m}{\partial \dot{
z}_i(t)}$, one would obtain after extremization in $\mu(t)$ the usual
answer for the Hamiltonian \be H_0=\sqrt{p^2_i+m^2}\ee

In case of zero mass, $ m=0$, one would obtain for the free gluon
without spin  from (A2.21) the free hamiltonian $\sqrt{\vec p^2}$.

The NP interaction in the two-gluon system is given by (4), where the term
$\int B_\mu dz_\mu$ generates the adjoint string, equation (5), and
after introduction of another einbein function $\nu(t)$, as it is
done in [14] one obtains the Hamiltonian (6).

The latter describes the straight-line adjoint string connecting two gluons,
which can rotate and change its length.

Contribution of NP spin terms is considered in Appendix 3.\\[1cm]

{\Large \bf  Appendix 3} \\

{\bf
 Nonperturbative spin--splitting terms}

\setcounter{equation}{0} \def\theequation{A3.\arabic{equation}}
 \vspace{1cm}

Introducing the spin matrix of the gluon as in (22) and using
(A2.18), one can rewrite the $\hat F$ terms in the exponent of (4)
as
\be
 2g\int
(SF(t))\frac{dt}{2\mu(t)}-2g\int(S'F(t'))\frac{dt'}{2\mu(t')}
\ee
where $(SF)\equiv S_iB_i+\tilde S_i E_i\equiv S_{\mu\nu}F_{\mu\nu}$.
Note that the contribution of the second (primed) gluon to (A3.1) has
another sign as compared to the first gluon. This is the consequence
of the fact, that color and time ordering of operators $B, F$ and
$B', F'$ are opposite in the closed loop $W_F$. One must use
therefore in $W_F$ the transposed  operators for the second (or the
first) gluon and write $B^ {\prime T}=-B'$.

To calculate the average of the exponential (4), one can use the following
trick: using the nonabelian Stokes theorem, and cluster expansion
in the Gaussian approximation one rewrites Wilson loop integral as
\be
\langle W_{adj}\rangle = \langle P_F expig \int\limits_s
F_{\mu\nu}(u)d\sigma_{\mu\nu}(u)\rangle =
exp[-\frac{g^2}{2}\int\limits_s\int\limits_s\langle
F_{\mu\nu}F_{\lambda\sigma}\rangle
d\sigma_{\mu\nu}d\sigma_{\lambda\sigma}]
\ee
Then (4) can be rewritten as
\be
\langle W_F\rangle=tr
exp[-2i\int\limits^T_0
\frac{dt}{2\mu(t)}S_{\mu\nu}\frac{\delta}{\delta\sigma_{\mu\nu}(u)}
+2i\int\limits^T_0\frac{dt'}{2\mu(t')}S'_{\mu\nu}
\frac{\delta}{\delta\sigma_{\mu\nu}(u)}]\langle W_{adj}\rangle
\ee

Evaluating derivatives, one arrives at the expression, based on the
Gaussian approximation
\be
 \langle W_F\rangle=tr
exp[-\frac{g^2}{2}\int\int[d\sigma_{\mu\nu}(u)-2i S_{\mu\nu}
\frac{dt}{2\mu(t)}][d\sigma_{\lambda\sigma}(u')+2i
S'_{\lambda\sigma}\frac{dt'}{2\mu(t')}] \langle
F_{\mu\nu}(u)F_{\lambda\sigma}(u')\rangle
\ee
One can add in the exponent of (A3.2) all higher correlators in
spin--independent terms and thus
restore the area law (5) with $\sigma_{adj}$
exact (i.e. beyond Gaussian approximation).
 For spin--dependent terms higher
correlators bring about higher powers of $S,S'$.

Since spin--dependent terms are relatively small corrections, it is
legitimate to keep for them lowest, i.e. Gaussian approximation and write
\be
\langle W_F\rangle\approx Z trexp (-\sigma_{adj} S_{min}) exp (N_1+
N_2+N_{12})\ee
where notations used are
\be
N_1=ig^2\int\int
 d\sigma_{\lambda\sigma}(u)\cdot S_{\mu\nu}\frac{dt}{2\mu(t)}
\langle F_{\mu\nu}(u)F_{\lambda\sigma}(\omega(t))\rangle
\ee
\be
N_{12}=2g^2\int\int
 \frac{dt}{2\mu(t)} \frac{dt'}{2\mu(t')}
S_{\mu\nu}S'_{\lambda\sigma}\langle
F_{\mu\nu}(u(t))F_{\lambda\sigma}(u'(t'))\rangle\ee
and $N_2$ is obtained from $N_1$ by an exchange $t\to t'$.

The transformations in (A3.6),(A3.7) into the spin--orbit,
spin--spin and tensor terms in (24) are the same as in the
corresponding heavy quarkonium expressions given in [25], which are
similar to (A3.6),(A3.7) modulo numerical coefficients and
different gluon spin factors.

The field correlators $\langle F F\rangle$ enter the final expression via
potentials $V_i(r),i=1,2,3,4$ which are the same as for heavy quarkonia and
given in [25], with the replacement $C_2(f)\to C_2(adj)$. If one
introduces two scalar functions $ D$ and $D_1$ as in [23], one can
write,
\be
\frac{1}{R}\frac{dV_1}{dR}=
-\int\limits^\infty_{-\infty}
dv\int\limits^R_0\frac{d\lambda}{R}(1-\frac{\lambda}{R})D(\lambda, v),
\ee
\be
\frac{1}{R}\frac{dV_2}{dR}=
\int\limits^\infty_{-\infty}dv
\int\limits^R_0\frac{\lambda d\lambda}{R^2}[D(\lambda,v)+D_1(\lambda,v)
+\lambda^2\frac{\partial
D_1}{\partial\lambda^2}]\ee
\be
V_3=- \int\limits^\infty_{-\infty}dv R^2 \frac{\partial D_1(R,v)}{\partial
R^2}\ee
\be
V_4=\int\limits^\infty_{-\infty}dv[3D(R,v)+3D_1(R,v)+R^2
\frac{\partial D_1}{\partial R^2}]\ee

Note that $D=\frac{C_2(adj)}{C_2(f)}D^f$, and the same relation for
$D_1$, where $D^f$, $D_1^f$ refer to the fundamental representation.
The normalization of $D$ can be obtained from the relation
\be
\sigma^{(2)}_{adj}=\frac{1}{2}\int d^2x D(x)\ee
where the subscript (2) in $\sigma_{adj}$ denotes the lowest (quadratic)
correlator contribution to the string tension. As one can argue, the accuracy
of this quadratic approximation is around 10\% [23].

Taking asymptotically large R in (A3.8), and using (A3.12) one
obtains the asymptotics (25), given in the main text.

To evaluate NP spin--orbit splitting, we must estimate the matrix element
$\langle V'_1+V'_2\rangle$, i.e. some integrals with $D,D_1$. The latter have
been measured on the lattice [27] and found to be of exponential form
\be
D(x)=D(0)exp(-x\delta), D_1(x)\ll D(x)
\ee
with $\delta\approx 1 GeV$. For an estimate of
$\langle V'_1+V'_2\rangle$ we
neglect $D_1$, and calculate
 \be
 \langle V'_1+V'_2\rangle
=\frac{2\sigma_{adj}}{\pi}(2[-J_1(x)
+\frac{2-x^2K_2(x)}{x}]+J_2(x))\ee
where $J_n(x), K_n(x)$ are Bessel and McDonald functions
respectively and $x=\delta r$.

>From (A3.14) one can see that asymptotic behaviour (25) is obtained
only for large $r>7\delta^{-1}\sim 1.5 fm$. Therefore the average of
$\Delta H(Thomas)$ with the square of the glueball eigenfunction is
considerably reduced as compared to the average of the asymptotics
(25), and the resulting NP spin--orbit term given in Table 7 is
smaller than the corresponding perturbative term, and the ordering of
the levels is due to the perturbative part of spin--dependent
forces.\\[1cm]

{\Large \bf  Appendix 4} \\

{\bf
 Derivation of the relation for the matrix element $\langle
r^{-3}\rangle$.}

\setcounter{equation}{0} \def\theequation{A4.\arabic{equation}}
 \vspace{1cm}

Writing solution of the Hamiltonian  $H'_0$,  Eq.(8) in the form
\be
\psi_n(r)=\frac{y_n(r)}{r}Y_{lm}, y_n(r)\sim r^{l+1}, r\to 0
\ee
one can rewrite equation as
$$
y^{\prime\prime}_n=[2\tilde
\mu(V(r)-E_n)+\frac{L(L+1)}{r^2}]y_n(r)\eqno(A 4.2) $$

We shall use the procedure suggested in
the second of ref. [29].
Multiplying both sides of (A4.2) with $\frac{y'_n(r)}{4\pi r^2}$ and
integrating over $d^3r$, one obtains
$$\int d^3r \frac{y''_n y'_n}{4\pi r^2}=-\frac{1}{2}[y'_n(0)]^2=
\int\limits^\infty_0dr
[2\tilde \mu(V(r)-E_n)+\frac{L(L+1)}{r^2}]\frac{(y^2_n)}{2}=
$$
\be
=-\frac{1}{2}\int\limits^\infty_0 y^2_n(r)dr
[2\tilde \mu V'(r)-2\frac{L(L+1)}{r^3}]\ee

Taking into account (A4.1) one has from two results:

For $L=0$ one obtains the well-known relation [29]
\be
|\psi_n(0)|^2=\frac{\tilde \mu}{2\pi}\langle V'(r)\rangle\ee

For $L>0$ one has instead,
\be
L(L+1)\langle\frac{1}{r^3}\rangle=\tilde \mu\langle
V'(r)\rangle\ee

In our case (Eq.(8)) $\mu_0=2\tilde \mu, V'(r)=\sigma_{adj}$.
Note that both l.h.s in (A4.4) and (A4.5) do not depend on the radial
quantum number $n$.\\[1cm]

{\Large \bf  Appendix 5} \\

{\bf
 Mixing of glueball and $f,f'$ trajectories.}

\setcounter{equation}{0} \def\theequation{A5.\arabic{equation}}
 \vspace{1cm}

One can start with the scattering amplitude of hadron $a$ on hadron $
b$ in the Regge-pole approximation
\be
 T^{(ab)}=\sum_{i,k}g^{(aa)}_i
T_{ik}g^{(bb)}_k\ee
 where $(i,k)=1,2,3$ refer to the bare
Regge trajectories
\be
 j=\bar\alpha_i(t), i=1,2,3\ee
  and the
matrix $T_{ik}$ has the form \be
 T_{ik}=(j-\hat\alpha(t)-\hat
g(t))^{-1}_{ik}\ee
 with the notations
 \be
(\hat\alpha(t))_{ik}=\bar\alpha_i(t)\cdot \delta_{ik},(\hat
g(t))_{ik}=g_{ik}(t),g_{ll}=0\ee

The nondiagonal matrix
 $\hat g(t)$ describes mixing of Regge trajectories. In
what follows we consider three bare trajectories: $\bar\alpha_1(t)$ is the
glueball trajectory calculated in the paper in
Section 6. We approximate it in the region $0\leq t\leq 6 GeV^2$ by
linear form
 \be
\bar\alpha_1(t)=\bar\alpha_1(0)+\bar\alpha'_1(0)\cdot
 t=0.7+0.246t
 \ee
 where the value for
$\bar\alpha'_1(0)
$
is chosen in such a way as to reproduce the
first glueball $2^{++}$ state at $M=2.3 GeV$
(Table 5).

The bare $f$ and $f'$ trajectories are denoted as $\bar\alpha_2(t)$
and $\bar\alpha_3(t)$ respectively and taken in the form
\be
\bar\alpha_2(t)=0.55+0.89t;~~ \bar\alpha_3(t)=0.25+0.83t\ee

The mixing matrix $g_{ik}(t)$ is not known theoretically; as was discussed in
section 6, condition of planarity [34-36] requires $g_{ik}(t)$ to
fall off at large positive $t$, therefore we assume for it the form
\be
g_{ik}(t) =\frac{g^{(0)}_{ik}}{1+(t/\lambda^2)^k}\ee
and for explicit calculations
in the region $t>0$ we use $k=1 $ and $\lambda^2=\frac{2}{3}GeV^2$.

To find shifted Regge poles in $T$, one can rewrite (A5.3) as
\be
T_{ik}=\frac{t_{ik}}{det(j-\hat\alpha(t)-\hat g(t))}\ee
where $t_{ik}$ are minors of $\hat T$. The roots of determinant in
(A5.8) are given by cubic equation
\be
j^3-j^2\sum\bar\alpha_i+j(\sum\limits_{i\ne
k}\bar\alpha_i\bar
\alpha_k-g^2_{ik})
-\bar\alpha_1\bar\alpha_2
\bar\alpha_3+\sum\limits_{i\ne k\ne l}
g^2_{ik}\bar\alpha_l-2g_{12}g_{13}g_{23}=0\ee

We denote three roots of (A5.9) by
\be
j=\alpha_i(t),i=1,2,3\ee

Let us start with $t=0$. Assuming for $\bar\alpha_i(0)$ the values in
(A5.5), (A5.6)  and  the following values  for
$g^{(0)}_{ik}$
\footnote{The value of $g_{12}^{(0)}$ can be estimated from the model
of  $f$--dominance and experimental data on residues of the Pomeron
and $f$--poles,  $g^{(0)}_{12}=0.3\div 0.5$. The $g_{13}^{(0)}$
coupling  is approximately $\sim 0.5 g^{(0)}_{12}$, while
$g^{(0)}_{23}\approx g^{(0)}_{12}g^{(0)}_{13}$.},
\be (g^{(0)}_{12})^2=0.16;~~
(g^{(0)}_{13})^2=0.08;~~g^{2}_{13}=0.01
\ee
we obtain the intercepts of mixed trajectories
\be
\alpha_1(0)=1.2;~~\alpha_3(0)=0.075; ~~\alpha_2(0)=0.225\ee

Thus we obtain a  realistic
intercept of the Pomeron, corresponding to the bare Pomeron intercept
observed in DIS' at small $x$ [20].
Note however, that theoretical uncertainty in $g^{(0)}_{ik}$ and in
the Pomeron intercept are rather larger ($\sim 0.1)$.

Resulting trajectories are depicted in Fig.1.

Comparing with bare trajectories $\bar\alpha_i(t)$ in Fig.1 one can
see that the role and ordering of trajectories is changed as compared
to bare ones when one goes from large $t$ to the small $t$ region.
 This property is very general and is not related to a particular
 choice of $g_{ik}(t)$.

It is also of interest to define the coupling of new Regge poles to the
hadrons $a,b$ and  to probe in this way the quark and gluon contents
of the poles. To this end we express the matrix $T_{ik}$ as \be
T_{ik}=O_{in}\lambda_n O^+_{nk}\ee
where the diagonal matrix $\hat\lambda$ is
\be
\hat\lambda=
\left(\begin{array}{ccc}
\frac{1}{j-\alpha_1(t)}&0&0\\
0&\frac{1}{j-\alpha_2(t)}&0\\
0&0&\frac{1}{j-\alpha_3(t)}\\
\end{array}\right)\ee
and find matrix elements $O_{ik}$ from the systems of equations
\be
(\alpha_k-\bar\alpha_1)O_{ik}+g_{12}O_{2k}+g_{13}O_{3k}=0\ee
\be
g_{12}O_{1k}+(\alpha_k-\bar\alpha_2)O_{2k}+g_{23}O_{3k}=0\ee
\be
g_{13}O_{1k}+g_{23}O_{2k}+(\alpha_k-\bar\alpha_3)O_{3k}=0\ee

Equations (A5.17)-(A5.19) for $k=1,2,3$ and normalization condition
\be
|O_{k1}|^2+|O_{k2}|^2+|O_{k3}|^2=1\ee
define $O_{ik}$ up to a common phase.

Physcially $|O_{ki}|^2=|O_{ik}|^2$ gives a probability of finding original
pole $\bar\alpha_i$ in the new pole $k$.

Since original indices $i$ refer to the glueball trajectory $(i=1)$,
$u\bar u+d\bar d$  $f$--trajectory $(i=2)$ and $s\bar s$ trajectory
$(i=3)$, one can define in this way the percentage of the
corresponding components in the new trajectory.

 \newpage

\begin{center}
{\bf Table 1}\\

Effective mass eigenvalues $\mu_0(n,l)$ (in GeV for $\sigma_f=0.18 GeV^2$)
  obtained from Eq.(8), $\mu_0=\sqrt{\sigma_{adj}}(\frac{a(n)}{3})^{3/4}$-
  upper entry,and eigenvalues of reduced equation a(n)--lower entry.
  \vspace{1cm}

\begin{tabular}{|l|l|l|l|l|} \hline
L$\setminus$n&0&1&2&3  \\ \hline
0&0.528&0.803&1.005&1.174\\
&2.3381&4.0879&5.520&6.786\\ \hline
1&0.693&0.917& & \\ &3.3613&4.8845& & \\  \hline
2&0.826&1.020& &  \\ &4.2482&5.6297& &  \\ \hline
  \end{tabular}
\vspace{1.5cm}

{\bf Table 2}\\
The eigenvalues (in GeV) of relativistic Hamiltonian for $L=0$
\vspace{1cm}

 \begin{tabular}{|l|l|l|l|l|l|l|} \hline
 $n$&0&1&2&3&4&5\\ \hline
 $M_n$&2.01&2.99&3.75&4.37&4.92&5.41\\\hline
 \end{tabular}
\vspace{1.5cm}

{\bf Table 3}\\
 The eigenvalues (in GeV) of rotating string Hamiltonian (6) for
$L>0$.
 \vspace{1cm}

 \begin{tabular}{|ll|l|l|l|l|l|} \hline
 &$L$&1&2&3&4&5\\
 $n$&&&&&& \\ \hline
 0&&2.65&3.13&3.53&3.88&4.206\\ \hline
 1&&3.645&4.03&4.366& 4.67&4.95\\     \hline
 2&&4.40&4.737&5.04&5.31&5.56\\             \hline
 3&&5.02&5.34&5.62&5.87&6.10\\\hline
 4&&5.58&5.87&6.13&6.37&6.59\\      \hline
 5&&6.09&6.36&6.60&6.82&7.03\\              \hline
 \end{tabular}
\vspace{1.5cm}

{\bf Table 4}\\

Spin averaged glueball masses $M_G/\sqrt{\sigma_f}$

\vspace{1cm}

 \begin{tabular}{|l|l|l|l|l|} \hline
\multicolumn{2}{|c|}{ Quantum}& This& \multicolumn{2}{c|}{Lattice data}\\
\cline{4-5}
\multicolumn{2}{|c|}{ numbers}&work& ref. [3]&ref. [1]\\\hline
 &$L=0,n_r=0$&4.68&4.66$\pm$0.14&4.55$\pm$0.23\\   \cline{2-5}
 2 gluon&$L=1,n_r=0$&6.0 &6.36$\pm$0.6 &6.1 $\pm$0.38 \\\cline{2-5}
 states&$L=0,n_r=1$&7.0 &6.68$\pm$0.6 &6.56$\pm$0.55\\\cline{2-5}
 &$L=2,n_r=0$&7.0 &9.0 $\pm$0.7(3$^{++}$) &7.7 $\pm 0.4(3^{++})$ \\
 \cline{2-5}
 &$L=1,n_r=1$&8.0 &  &7.94 $\pm 0.48$ \\ \hline
 3 gluon&K=0&7.61&&8.19$\pm$0.48\\
 state&&&&\\\hline
   \end{tabular}
\vspace{1.5cm}

{\bf Table 5}\\
{Masses of glueballs with $L=0,1,2$ and $n=0,1$, $\sigma_f=0.18
GeV^2$}\\
\vspace{1cm}

\begin{tabular}{|c|c|c|c|c|c|c|c|c|}\hline
$J^{PC}$
 & \multicolumn {2}{|c}{$0^{++}$}
 & \multicolumn {2}{|c}{$2^{++}$} &
\multicolumn {2}{|c}{$0^{-+}$}
 & \multicolumn {2}{|c|}{$2^{-+}$}
\\\hline
n & 0 & 1 & 0 & 1 & 0 & 1& 0 & 1\\ \hline
M(GeV)
 & 1.4 & 2.4 & 2.3 & 3.3 & 2.52 & 3.55 & 2.70 & 3.7\\ \hline
\end{tabular}
\vspace{1cm}

\begin{tabular}{|c|c|c|c|c|c|c|c|c|}\hline
$J^{PC}$
  & \multicolumn
{1}{|c}{$2^{++}$} & \multicolumn{1}{|c}{$0^{++}$} & \multicolumn
{1}{|c}{$1^{++}$} &
 \multicolumn {1}{|c}{$3^{++}$}
 &\multicolumn{1}{|c}{$4^{++}$}
& \multicolumn {1}{|c}{$3^{--}$}
& \multicolumn {1}{|c}{$2^{--}$}
& \multicolumn {1}{|c|}{$1^{--}$}
 \\ \hline
n & 0 & 0 & 0 & 0 & 0 & 0& 0 &0\\ \hline
M(GeV)
 & 3.13; 3.11 &
3.06 & 3.07 & 3.14 & 3.16&3.51&3.23&3.04  \\ \hline
\end{tabular}

\newpage

{\bf Table 6}\\

Comparison of predicted glueball masses with lattice data (for
$\sigma_f=0.238 GeV^2  $

\vspace{0.5cm}

 \begin{tabular}{|l|l|l|l|} \hline
$J^{PC}$& M(GeV)&  \multicolumn{2}{c|}{Lattice data}\\\cline{3-4}
 &This work& ref. [1]&ref. [3]\\\hline

$0^{++}$&1.58&1.73$\pm$0.13&1.74$\pm$0.05\\
$0^{++*}$&2.71&2.67$\pm$0.31&3.14$\pm$0.10\\
$2^{++}$&2.59&2.40$\pm$0.15&2.47$\pm$0.08\\
$2^{++*}$&3.73&3.29$\pm$0.16&3.21$\pm$0.35\\
$0^{-+}$&2.56&2.59$\pm$0.17&2.37$\pm$0.27\\
$0^{-+*}$&3.77& 3.64$\pm$0.24&          \\
$2^{-+}$&3.03&3.1$\pm$0.18&3.37$\pm$0.31\\
$2^{-+*}$&4.15& 3.89$\pm$0.23&          \\
$3^{++}$&3.58&3.69$\pm$0.22&4.3$\pm$0.34\\
$1^{--}$&3.49& 3.85$\pm$0.24&          \\
$2^{--}$&3.71& 3.93$\pm$0.23&          \\
$3^{--}$&4.03& 4.13$\pm$0.29&          \\\hline
  \end{tabular}
\newpage

{\bf Table 7\\ }

  Spin-orbit and tensor corrections to two-gluon glueball masses (the upper
entries for $n=0$ and lower for $n=1$)$\sigma_f=0.18 GeV^2$
\vspace {0.5cm}

\begin{tabular}{|l|l|l|l|l|l|l|l|} \hline
 & \multicolumn{2}{|c|}{L=S=1} & \multicolumn{5}{|c|}{L=S=2} \\
 \hline $J^{PC}$ & $0^{-+}$ & $2^{-+}$ & $0^{++}$ & $1^{++}$ &
$2^{++}$ & $3^{++}$ & $4^{++}$  \\
\hline $M_{SL}^{(pert)}\vec S \vec L$ & -0.197 & -0.0985 & - 0.1656 &
- 0.138 & -0.083 & 0 & 0.110 \\ &-0.148&
0.074&-0.128&-0.107&-0.064&0&0.085\\ \hline
$M_T^{(pert)}$ & -0.263 &
0.0263&-0.072 &-0.036& 0.015 & 0.041 & 0.020\\
& -0.198
&-0.02&-0.056&-0.028&0.0116&0.032&0.0155\\ \hline $\Delta
M^{tot}_{pert}$&-0.46&+0.072&-0.238&-0.174&-0.068&0.041&0.13\\
&-0.347&+0.054&-0.185&-0.135&-0.053&0.032&0.101\\ \hline
$\Delta M^{Thom}\cdot \vec S\vec
L$&0.082&-0.041&0.216&0.18&0.108&0&-0.144\\
&0.05&-0.025&0.138&0.115&0.07&0&-0.092\\ \hline
$ \Delta M_{tot}$&-0.38&+0.031&-0.022&+0.006&0.04&0.041&-0.014\\
&-0.3&+0.029&-0.047&-0.02&0.017&0.032&0.009\\ \hline
$<S_{12}>$&-2&-1/5&-2&-1&3/7&  8/7&-4/7\\\hline
$<\vec L\vec S>$&-2&+1&-6&-5&-3&0&4\\\hline
 \end{tabular}
\vspace{1cm}

{\bf Table 8}\\
Effect of inclusion of the Coulomb interactions on glueball masses
and Regge paramerers,
$M(\alpha_s, L=0,1,2), \sigma_f=0.18 GeV^2$
\vspace{0.5cm}

\vspace{1cm}
 \begin{tabular}{|l|l|l|l|l|} \hline
 $\alpha_s$&0&0.2&0.3&0.39  \\ \hline
 $M_0(L=0)$&2.11&1.776&1.587&1.390\\ \hline
 $M_0(L=1)$&2.77&2.56&2.45&2.36  \\ \hline
 $M_0(L=2)$&3.30&3.14&3.05&2.97  \\ \hline
$\alpha'_G(0)$&0.31&0.298&0.294&0.290  \\ \hline
$\alpha_G(0)$&0.617&1.06&1.259&1.44  \\  \hline
\end{tabular}

\newpage

{\bf Table 9}\\

 Two--gluon glueball operators

\vspace{0.5cm}

 \begin{tabular}{|c|l|l|l|l|l|} \hline
 Dimension& $J^{PC}$&$\Psi^{(in)},
 \Psi^{(out)}$&$\Psi^{(in),(out)}$ in BPT& $L$&BPT
 Dimension\\ \hline
 4&$0^{++}$& $tr (E_iE_i)$& $tr(a_ia_i)$&0&2\\
 4&$2^{++}$& $symm_{ik}~tr (E_iE_k)$& $symm_{(ik)}~tr(a_ia_i)$&0&2\\
 4&$0^{-+}$& $tr (E_iB_i)$& $tr(a_i({\bf D}\times {\bf a})_i)$&1&3\\
 4&$1^{-+}$& $tr ({\bf E}\times {\bf B})$& $tr({\bf a}\times ({\bf
 D}\times {\bf a}))$&1&3\\
 4&$2^{-+}$& $symm_{(ik)}~tr ( E_iB_k)$& $ symm_{(ik)}~tr(
a_i ({\bf D}\times {\bf a})_k)$&1&3\\
 4&$0^{++}$& $tr ( B_iB_i)$& $ tr(
 ( ({\bf D}\times {\bf a})\cdot
  ({\bf D}\times {\bf a}))$&2&4\\
 4&$2^{++}$& $symm_{(ik)}~tr ( B_iB_k)$& $ symm_{(ik)}~tr(
({\bf D}\times {\bf a})_i)
({\bf D}\times {\bf a})_k)$&2&4\\
 6&$3^{++}$& $symm_{(klm)}~tr ( D_4E_kD_lB_m+D_4B_mD_lE_k)$& $
 symm_{(ikl)}~tr( ({\bf D}\times {\bf a})_i) D_ka_l)
$&2&4\\
 6&$4^{++}$& $symm_{(ilkm)}~tr ( D_iE_kD_lE_m)$& $
 symm_{(iklm)}~tr( D_ia_kD_la_m)$&2&4 \\ \hline
 \end{tabular}

\vspace{1cm}

{\bf Table 10}\\

 Three--gluon glueball operators

\vspace{0.5cm}

 \begin{tabular}{|c|l|l|l|l|l|} \hline
 Dimension& $J^{PC}$&$\Psi^{(in)},
 \Psi^{(out)}$&$\Psi^{(in),(out)}$ in BPT& $L$&BPT
 Dimension\\ \hline
 6&$1^{+-}$& $tr (\{E_k,E_l\} B_l)$& $tr(\{a_ka_l\}({\bf D}\times
 {\bf a})_l)$&1&4\\
 6&$3^{+-}$& $symm_{klm}~tr (\{E_k,E_l\} B_m)$&
 $symm tr(\{a_k,a_l\}({\bf D}\times {\bf a})_m\}$&1&4\\
 6&$2^{+-}$&
 $symm_{(kn)} e_{nlm} tr (\{E_k, E_l\}B_m)$& $ symm_{(kn)}
 e_{nlm} tr(\{a_k, a_e\}({\bf D}\times {\bf a})_m)$&1&4\\
 6&$1^{--}$& $tr ( E_iE_kE_k)$& $tr(a_ia_ka_k)$&0&3\\
 6&$3^{--}$&
 $symm_{(klm)}~tr ( E_kE_lE_m)$& $ symm_{(klm)}~tr( a_ka_la_m)
 $&0&3\\
  \hline
  \end{tabular}

 \end{center}

             \newpage

             \begin{center}

{\bf  Table captions}

             \end{center}

{\bf Table 1}\\

Effective mass eigenvalues $\mu_0(n,l)$ (in GeV for $\sigma_f=0.18 GeV^2$)
  obtained from Eq.(8), $\mu_0=\sqrt{\sigma_{adj}}(\frac{a(n)}{3})^{3/4}$-
  upper entry,and eigenvalues of reduced equation a(n)--lower entry.
  \vspace{1cm}

{\bf Table 2}\\
The eigenvalues (in GeV) of relativistic Hamiltonian for $L=0$
\vspace{1cm}

{\bf Table 3}\\
 The eigenvalues (in GeV) of rotating string Hamiltonian (6) for
$L>0$.
 \vspace{1cm}

{\bf Table 4}\\

Spin averaged glueball masses $M_G/\sqrt{\sigma_f}$

\vspace{1cm}

{\bf Table 5}\\
{Masses of glueballs with $L=0,1,2$ and $n=0,1$, $\sigma_f=0.18
GeV^2$}\\
\vspace{1cm}

{\bf Table 6}\\

Comparison of predicted glueball masses with lattice data (for
$\sigma_f=0.238 GeV^2  $

\vspace{0.5cm}

{\bf Table 7\\ }

  Spin-orbit and tensor corrections to two-gluon glueball masses (the upper
entries for $n=0$ and lower for $n=1$)$\sigma_f=0.18 GeV^2$
\vspace {0.5cm}

{\bf Table 8}\\
Effect of inclusion of the Coulomb interactions on glueball masses
and Regge paramerers,
$M(\alpha_s, L=0,1,2), \sigma_f=0.18 GeV^2$
\vspace{0.5cm}

{\bf Table 9}\\

 Two--gluon glueball operators

\vspace{0.5cm}

{\bf Table 10}\\

 Three--gluon glueball operators

\newpage
\begin{figure}[t]
\epsfxsize=19cm
\epsfbox{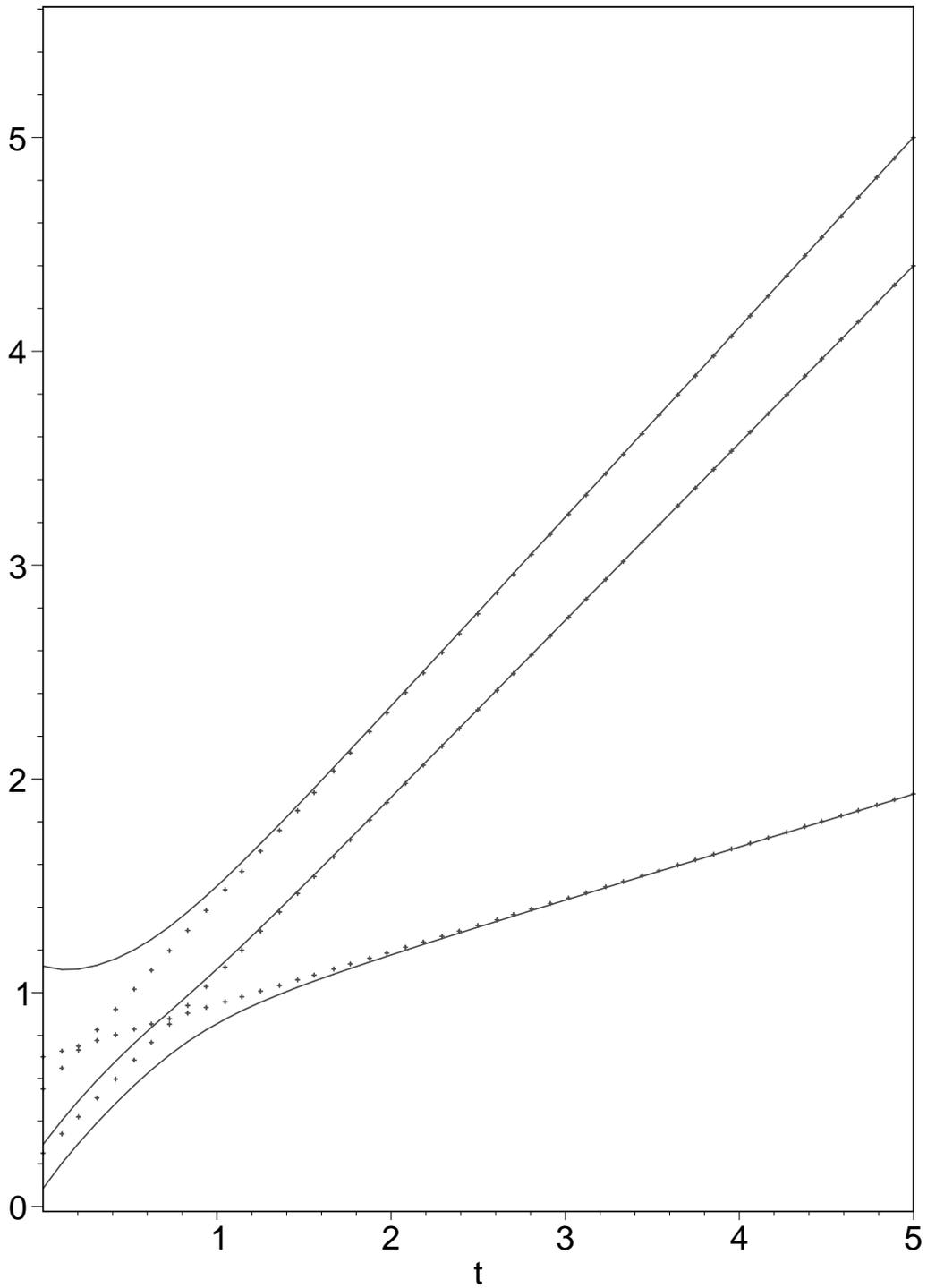}
\caption{
Glueball, $f$ and $f'$ Regge trajectories as functions of
$t$--channel energy squared $t$ (in $GeV^2$). Dotted lines --- bare
trajectories, solid lines --- trajectories with the coupling
$g_{ik}(t)$ taken into account in the form (A5.7), with parameters
$k=1, \lambda^2=\frac{2}{3} GeV^2$.}
\end{figure}
\newpage

\begin{figure}[t]
\epsfxsize=19cm
\epsfbox{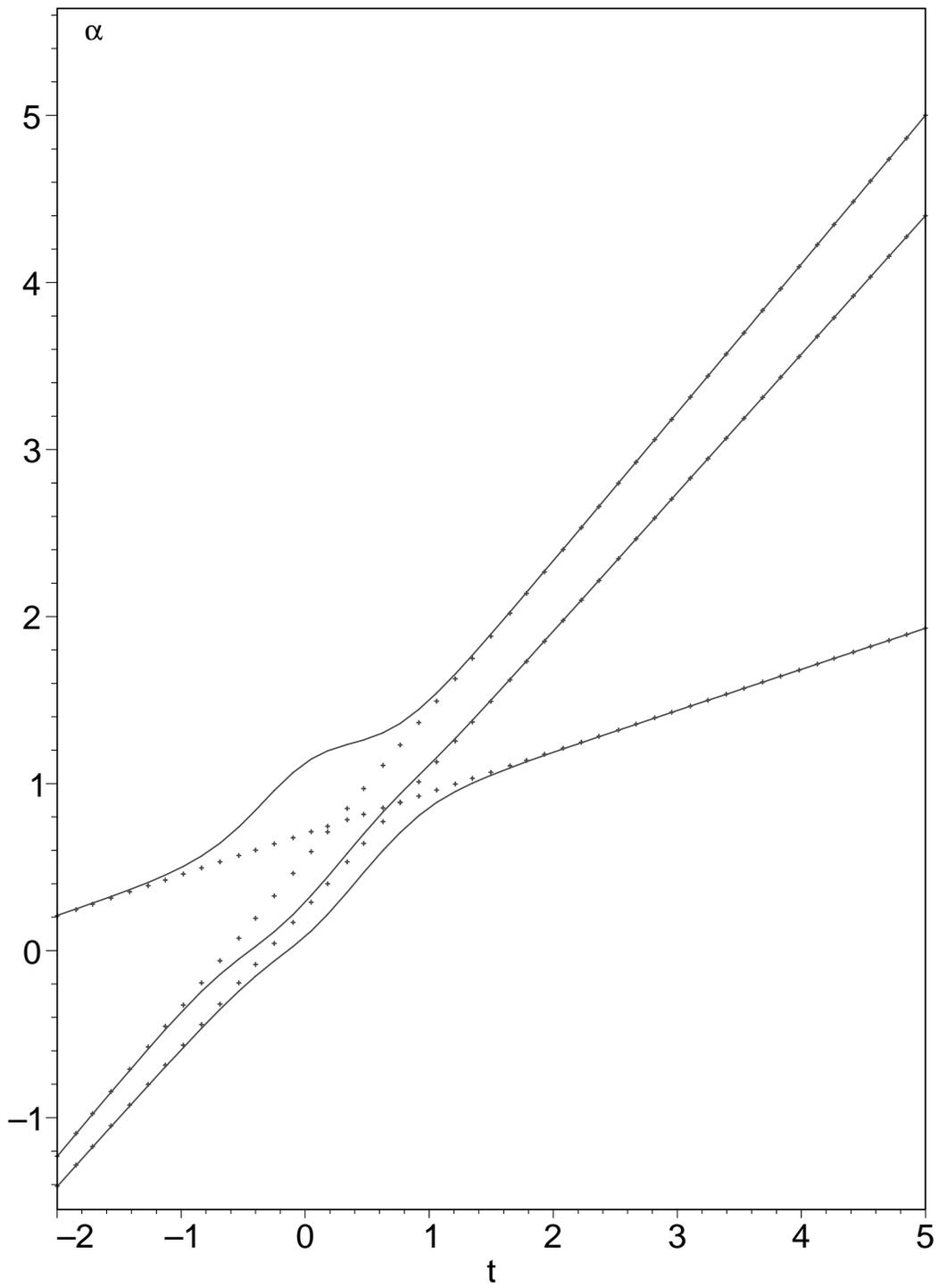}
\caption{The same as in Fig.1 but for the coupling parameters $k=2,
\lambda^2=\frac23 GeV^2$.
}
\end{figure}

\end{document}